\documentclass[a4paper,11pt]{article}
\usepackage{jheppub} % for details on the use of the package, please see the JINST-author-manual
\usepackage{lineno}
\usepackage{feynmp-auto}
\usepackage{subfigure}
\usepackage{subcaption}
\usepackage{mathabx}
\usepackage{nicematrix}
\usepackage{cleveref}
\usepackage{amsmath}
\usepackage[normalem]{ulem}
\usepackage{comment}
\usepackage{tikz-feynman}
\usepackage{snapshot}
\renewcommand\({\left(}
\renewcommand\){\right)}

\newcommand{\nn}{\nonumber}

\newcommand\eps{\epsilon}

\def\A{\mathcal{A}}
\def\L{\mathcal{L}}
\def\O{\mathcal{O}}
\let\vec\mathbf

\usepackage{subfiles} % Best loaded last in the preamble

\title{Triple Higgs boson production and electroweak phase transition in the two-real-singlet model
}

\author[1]{Osama Karkout,}
\author[2]{Andreas Papaefstathiou,}
\author[1,3]{Marieke Postma,}
\author[4,5]{Gilberto Tetlalmatzi-Xolocotzi,}
\author[6]{Jorinde van de Vis,}
\author[1]{Tristan du Pree}
\affiliation[1]{Nikhef, Theory Group, Science Park 105, 1098 XG Amsterdam, The Netherlands}
\affiliation[2]{Department of Physics, Kennesaw State University, 830 Polytechnic Lane, Marietta, GA 30060, USA}
\affiliation[3]{Institute for Mathematics, Astrophysics and Particle Physics, Radboud University Nijmegen, Heyendaalseweg 135, Nijmegen, the Netherlands}
\affiliation[4]{Universit\`e Paris-Saclay, CNRS/IN2P3, IJCLab, 91405 Orsay, France}
\affiliation[5]{Theoretische Physik 1, Center for Particle Physics Siegen (CPPS), Universit \"at Siegen, Walter-Flex-Str. 3, 57068 Siegen, Germany}
\affiliation[6]{Instituut-Lorentz for Theoretical Physics, Leiden University, Niels Bohrweg 2, 2333 CA Leiden, the Netherlands}
\emailAdd{o.karkout@nikhef.nl}
\emailAdd{apapaefs@kennesaw.edu}
\emailAdd{mpostma@nikhef.nl}
\emailAdd{gtx@physik.uni-siegen.de}
\emailAdd{vandevis@lorentz.leidenuniv.nl}
\emailAdd{tdupree@nikhef.nl}

\abstract{The production of three Higgs bosons at hadron colliders can be enhanced by a double-resonant effect in the $\mathbb{Z}_2$-symmetric two-real-singlet extension of the Standard Model, making it potentially observable in future LHC runs. The production rate is maximized for large scalar couplings, which prompts us to carefully reconsider the perturbativity constraints on the theory. This leads us to construct a new set of 140 benchmark points that have a triple Higgs boson production cross-section at least 100 times larger than the SM value.

Furthermore, we study the dynamics of the electroweak phase transition, both analytically at leading order, and numerically without the high-temperature expansion. Both analyses indicate that a first-order phase transition is incompatible with the requirement that both singlets have a non-zero vev in the present-day vacuum, as required by doubly-enhanced triple Higgs boson production. Allowing instead one of the singlets to remain at zero field value opens up the possibility of a first-order phase transition, while di-Higgs boson production can still be enhanced by a (single) resonance.
}

\begin{document}
\begin{flushright}

SI-HEP-2024-09\\

P3H-24-025\\
Nikhef 2022-025\\
MCNET-24-05

\end{flushright}

\maketitle
\flushbottom

\section{Introduction}\label{sec:intro}

The Higgs boson~\cite{Higgs:1964pj,Englert:1964et,Guralnik:1964eu} has been under intense experimental scrutiny since its discovery by the CERN ATLAS and CMS collaborations in 2012~\cite{ATLAS:2012yve,CMS:2012qbp}. Measurements of its properties, including its production and decay rates, are thus far consistent with the Standard Model (SM) Higgs boson~\cite{CMS:2022dwd,ATLAS:2024fkg}. Despite these results, there is no strict theoretical requirement for the Higgs sector to be minimal. In fact, there are several reasons to favor a non-minimal scalar sector. For example, the SM does provides no explanation for the nature of the non-luminous ``dark'' matter that permeates our universe, nor does it explain the flavor structure, nor the theoretically puzzling hierarchy problem. Another embarrassing problem for the SM is the unknown origin of the cosmic baryon asymmetry, also known as the matter-antimatter asymmetry. 

``Electroweak baryogenesis''~\cite{Kuzmin:1985mm, Shaposhnikov:1986jp, Shaposhnikov:1987tw, Cohen:1990py, Cohen:1993nk} is a promising scenario that can generate this asymmetry. It demands the electroweak phase transition to be first order, providing the necessary departure from thermodynamic equilibrium (one of the Sakharov conditions~\cite{Sakharov:1967dj}). Within the SM, the occurrence of a first-order phase transition (FOPT) requires the mass of the Higgs boson to lie below $\sim 70$~GeV~\cite{Bochkarev:1987wf,Kajantie:1995kf,Kajantie:1996qd,Csikor:1998eu}, which is evidently inconsistent with the experimental observation of the Higgs boson mass of $\sim 125$~GeV. Extended scalar sectors can accommodate an FOPT that is, moreover, sufficiently strong to preserve any baryon asymmetry produced. We note further that the gravitational wave signal from an FOPT at the electroweak scale is in the right frequency range to be potentially observable with the LISA gravitational wave observatory~\cite{Grojean:2006bp, Caprini:2018mtu,Caprini:2019egz,LISACosmologyWorkingGroup:2022jok}. 

The SM scalar sector can be extended by an arbitrary number of scalars~\cite{Chung:2012vg, Tenkanen:2022tly, Zuk:2022qwx}, but, arguably, the most minimal options are to add a complex or real singlet scalar field~\cite{Espinosa:1993bs, Benson:1993qx,Vergara:1996ub,Espinosa:2008kw,Barger:2007im,Espinosa:2011ax,Cline:2012hg,Curtin:2014jma,Katz:2014bha,Kanemura:2015fra,Kurup:2017dzf,Carena:2019une,Papaefstathiou:2020iag,Papaefstathiou:2021glr,Chen:2022vac,Azatov:2022tii}, two real singlet fields~\cite{Robens:2019kga,Papaefstathiou:2020lyp,Robens:2021zvr,Robens:2022lkn,Robens:2022hvu,Robens:2023pax,Robens:2023oyz}, or additional doublet fields~\cite{Land:1992sm,Cline:1996mga,Turok:1991uc,Dorsch:2013wja,Dorsch:2014qja,Basler:2016obg,Andersen:2017ika,Dorsch:2017nza,Bernon:2017jgv,Low:2020iua}. The additional fields could induce a rich set of phenomena at colliders. In particular, if they mix with the SM-like Higgs boson, their corresponding physical states could be produced singly and decay into fermions or gauge bosons, yielding a signal in multi-lepton or jet searches for heavy resonances.

In this work, we focus on the $\mathbb{Z}_2$-symmetric two real singlet model (TRSM). The corresponding phenomenology is qualitatively very similar
to that of the SM augmented with a single real scalar (xSM). For example, in both set-ups di-Higgs boson production can be resonant if an intermediate singlet state is produced on-shell, and rates are enhanced compared to the SM. Where the TRSM gives genuinely new physics and correspondingly novel collider phenomenology compared to the xSM is in triple Higgs production. The tree-level amplitude for this process involving both singlet states can be enhanced by a double resonance, while the enhancement in the xSM can come from at most a single resonance. 

For this reason, we focus on triple Higgs boson production, which is the most challenging of the multi-Higgs boson production final states contemplated by the experimental collaborations at the moment. At the CERN Large Hadron Collider (LHC) running at a 14 TeV center-of-mass, the SM cross section for this process is $\mathcal{O}(0.1)$~fb~\cite{Maltoni:2014eza}, rendering any prospects for detection dire, at any level of significance. However, it has been shown that triple Higgs boson production in the TRSM may be enhanced to a level sufficient for observation and exploration~\cite{Papaefstathiou:2020lyp}. This requires that some of the new quartic couplings in the potential are sufficiently large.

Triple Higgs boson production is the first attainable process that allows to directly probe the quartic Higgs self-coupling in the scalar potential. Its observation would provide a clear signal for an extended scalar sector, as well as hints on its intricate structure, and possible r\^ole in electroweak baryogenesis. In fact, one might expect that there exists a correlation between enhanced multi-Higgs boson production and a strong FOPT necessary for baryogenesis, as they both require sufficiently large scalar couplings. Indeed, an FOPT can only arise if the changes to the SM scalar potential are significant, and previous studies of the electroweak phase transition in, e.g.\ the xSM, shows a clear preference for large couplings \cite{Kotwal:2016tex, Alves:2018oct, Papaefstathiou:2021glr}.
This intriguing observation constitutes the main motivation for the present article.

Our goal here is to extend and explore further the parameter space for enhanced triple Higgs boson production, and cross-correlate this with the parameter space regions that give rise to a strong FOPT. In doing so, we strive to take all theoretical constraints, and the most recent experimental bounds, into account. Given the large quartic couplings, care should be taken that the theory is still perturbative, and under theoretical control, at least at the scale of the heaviest singlet state. The perturbativity bounds we derive are stronger than those previously employed in the literature \cite{Robens:2019kga,Papaefstathiou:2020lyp}. We thus provide a set of updated benchmark points for triple Higgs boson production in the $\mathbb{Z}_2$-symmetric TRSM, for which the calculations can be considered to be more robust.

We then study the possibility of an FOPT both in (1) the scenario with the phase transition induced by the temperature-dependence of the leading-order thermal mass corrections, as well as (2) a transition with a radiatively-generated barrier. Surprisingly, we find that the parameter space for large triple Higgs production boson and for an FOPT do not overlap. This result is intricately linked to the $\mathbb{Z}_2$-symmetric nature of the set-up and the choice of the vacuum today. Doubly-resonant triple Higgs boson production requires both additional scalars to mix with the Higgs field, and thus both to have a non-zero vacuum expectation value (vev) at zero temperature.  If we give up on this requirement, and allow for one or both singlets to have zero vev, an FOPT becomes possible. Like in the xSM, di-Higgs boson production enhancement is possible if only one added scalar has a non-zero vev at low temperature. Unlike the xSM however, the TRSM introduces the possibility of an FOPT resonant di-Higgs boson production.

The present article is organized as follows: in \cref{sec:model} we discuss the scalar sector model of the $\mathbb{Z}_2$-symmetric two real singlet extension of the SM, discuss the conditions for large triple Higgs boson production, and their implications for the perturbativety bound. \Cref{sec:scan} contains our exploration of the parameter space of the model, including a list of updated benchmark points relevant to triple Higgs boson production. In \cref{sec:thermal}, we review the thermal history of the model, and analyze the possibilities for an FOPT both in the single- and multi-step scenario. We conclude in \cref{sec:conclusions}. Several useful relations for the model are deferred to \cref{A:trmsrelations}, such as the constraints on the boundedness of the potential, the connection between the mass and flavor basis, and the renormalization group equations (RGEs).

%%%%%%%%%%%%%%%%%%%%%%%%%%%%%%%%%%%%%%%%%%%%%%%%%%%%%%%%%%%%%
%%%%%%%%%%%%%%%%%%%%%%%%%%%%%%%%%%%%%%%%%%%%%%%%%%%%%%%%%%%%%
%%%%%%%%%%%%%%%%%%%%%%%%%%%%%%%%%%%%%%%%%%%%%%%%%%%%%%%%%%%%%
\section{Triple Higgs production in the two real singlet model}\label{sec:model}

\subsection{The two real singlet model }
\label{sec:TRSM}

The two-real singlet model (TRSM) comprises of the SM augmented by two real scalar singlets. The Lagrangian of the scalar sector is then:
\begin{align}
    {\cal L}_{\rm scalar}&= |
    D_\mu \Phi_1|^2 + (\partial_\mu \Phi_2)^2 + (\partial_\mu \Phi_3)^2 -V(\Phi_i), \nn \\
V(\Phi_i) &=- \sum_i \mu^2_i (\Phi_i^{\dagger}\Phi_i ) + \sum_{i \leq j}
\lambda_{ij} (\Phi_i^{\dagger}\Phi_i )(\Phi_j^{\dagger}\Phi_j ),
\label{L_TRSM}
\end{align}
with $i = 1,2,3$, and where $\Phi_1$ the SM Higgs field and $\Phi_{2,3}$ the additional scalars.  Two discrete $\mathbb{Z}_2^{(\Phi_2)} \otimes \mathbb{Z}_2^{(\Phi_3)}$  symmetries, under which the respective singlets are odd, ensure that the Lagrangian is quadratic in the new fields.
$D_\mu$ denotes the covariant derivatives, including the conventional SM gauge contributions, and the $\partial_\mu$ partial derivatives. 

The amplitude for doubly-resonant enhanced triple Higgs production is only non-vanishing if all scalars mix with each other. We are thus interested in the set-up where all scalars obtain a vacuum expectation value and possess  $\mu_i^2>0$ for $i=1,2,3$. 
The fields are expanded around their corresponding vev (in unitary gauge):
\begin{equation}
    \Phi_1 = \begin{pmatrix} 0\\\frac{v_1+\varphi_1}{\sqrt{2}}\end{pmatrix}, \qquad
    \Phi_i = \frac{v_i + \varphi_i}{\sqrt{2}} \quad {\rm for} \quad i=2,3.
    \label{eq:fields}
\end{equation}

The mass matrix $\partial_{\phi_i}\partial_{\phi_j} V$ is off-diagonal and can be diagonalized by a $3\times 3$ unitary matrix $R$ which depends on three mixing angles -- see \cref{A:basis} for explicit expressions. The mass eigenstates, which we denote by $h_i$, and the interaction eigenstates $\phi_i$, are related via $ h_i = R_{ij} \phi_j$. The scalar potential in \cref{L_TRSM} depends on 9 parameters: three mass parameters $\mu_i$, and six quartic couplings $\lambda_{ij}$.  Instead of $\mu_i$ and $\lambda_{ij}$, we can instead use the three mass eigenstates $M_i$, three mixing angles $\theta_{ij}$, and three vevs $v_i$, as the independent parameters defining the model, with the explicit relations for these given in \cref{A:basis}. We identify the lightest mass eigenstate $M_1$, with the physical Higgs mass, with the eigenstates satisfying the ordering $M_1 < M_2 <M_3$. Two parameters are then fixed by observations: the Higgs vev $v_1\simeq 246\,$GeV and the measured physical Higgs boson mass $M_1 \simeq 125\,$GeV.  

We point out that if one of the singlets has zero vev $v_i=0$ for $i=2,3$, the corresponding field $\phi_i$ does not mix with the other fields. This scenario does not offer novel collider phenomenology compared to the xSM. However, as we will see in \cref{sec:thermal}, it does have an interesting effect on cosmology, as it allows for an FOPT.
\subsection{Resonant triple Higgs boson production}
 \label{sec:old_BM}

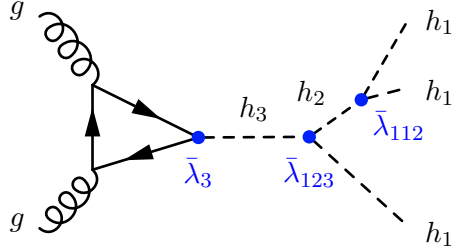
\begin{figure}[htp]
    \centering
    \setlength{\unitlength}{0.7cm}

\subfigure{
\begin{fmffile}{triangle2i2hhh}
    \begin{fmfgraph*}(7,4)
   \fmfstraight
    \fmfleft{i1,i2}
    \fmfright{o1,m,o2,o3}
    % gluons
    \fmf{gluon,tension=2}{i1,t1}
    \fmf{gluon,tension=2}{t2,i2}
    %\fmf{phantom,tension=0.4}{t1,o1}
    %\fmf{phantom,tension=0.4}{t2,o2}
    %\fmffreeze
    % top loop
    \fmf{fermion,tension=1}{t1,t2,t3,t1}
    \fmf{phantom,tension=1}{t3,m}
    \fmffreeze
    % Higgs boson
    \fmf{dashes,label=$h_3$,tension=2.5}{t3,h}
    \fmf{dashes,tension=1.2}{h,o1}
    \fmf{dashes,label=$h_2$,l.s=left,tension=3}{h,v1}
    \fmf{dashes,tension=2.2}{v1,o2}
    \fmf{dashes,tension=1.0}{v1,o3}
    \fmfv{decor.shape=circle,decor.filled=full,decor.size=4,f=(.0,,.13,,.98),
          l=$\color{blue}\bar{\lambda}_{112}$,l.d=6,l.a=-45}{v1}
    \fmfv{decor.shape=circle,decor.filled=full,decor.size=4,f=(.0,,.13,,.98),
          l=$\color{blue}\bar{\lambda}_{123}$,l.d=8,l.l=2,l.a=-90}{h}
    \fmfv{decor.shape=circle,decor.filled=full,decor.size=4,f=(.0,,.13,,.98),
          l=$\color{blue}\bar{\lambda}_{3}$,l.d=8,l.a=-90}{t3}
    \fmfv{l.a=-20,l=$h_1$}{o1}
    \fmfv{l.a=-20,l=$h_1$}{o2}
    \fmfv{l.a=-20,l=$h_1$}{o3}
    \fmfv{l.a=160,l=$g$}{i1}
    \fmfv{l.a=160,l=$g$}{i2}
    \end{fmfgraph*}
\end{fmffile}
}
\caption{Double-resonant triple SM-like Higgs boson ($h_1$) production in a model with two heavy scalars $h_3$ and $h_2$.}
    \label{fig:hhhdoubleres}
\end{figure}
 
With additional singlets, new diagrams with intermediate scalar states can contribute to triple Higgs boson production~\cite{Robens:2019kga}.
Although these amplitudes are suppressed by the small mixing angles, this can be overcome by resonance effects if the intermediate states are produced on-shell and above the threshold $M_3 > M_2 + M_1$ and $M_2 > 2M_1$. In the TRSM, double resonances can occur if both extra singlets are resonantly produced, in a contribution represented by the diagram shown in fig.~\ref{fig:hhhdoubleres}.  

To be concrete, consider the following triple Higgs boson production channels:
\begin{align}
    \A_1: \; pp \to h_3 \to h_2 h_1 \to h_1 h_1 h_1, \qquad
   \A^{(a)}_2: \; pp \to h_a  \to h_1 h_1 h_1, \quad a =1,2,3. 
\end{align}
The amplitudes for the above processes factorize into $h_3$ or $h_a$ production from a proton-proton collision, followed by the subsequent decay into three SM Higgs bosons. In the narrow width approximation for the intermediate $h_a$ propagators, the cross-section factorizes as well. The amplitude for initial $h_a$ production can be viewed as the production of the mostly SM Higgs-like interaction state
-- this rate is the same as in the  SM -- projected onto the mass eigenstates $h_a$: $\A_{pp\to h_a} = \kappa_{a}\ A^{\rm SM}_{pp\to h_1}$ \cite{Robens:2019kga},
where we define the scaling factors as
\begin{equation}
\kappa_a=R_{a1}.\label{eq:kappa}
\end{equation}
The second part of the amplitudes depends on the triple and quartic interactions of the mass eigenstates, which are defined as \cite{Robens:2019kga}
\begin{equation}
    \L = -\bar\lambda_{abc} h_a h_b h_c - \frac12 \bar\lambda_{aab} h_a^2 h_b- \frac1{3!} \bar\lambda_{aaab} h_a^3h_b +...\;,
\end{equation}
with 
\begin{align}
    \bar\lambda_{abc}&=(M_a^2+M_b^2+M_c^2)\sum_j \frac{R_{aj}R_{bj}R_{cj}}{v_j}, \nn \\
    \bar\lambda_{aaab}&=(3!) \sum_{ijk}
    \frac{M_k^2}{v_i v_j} R_{ki} R_{kj}(R_{ai}^2 R_{aj} R_{bj}+ R_{ai} R_{bi} R_{aj}^2),
\end{align}
and $R$ the mixing matrix of \cref{eq:Rmat}. The tree-level amplitudes can then be written as (up to symmetry factors)
\begin{align}
    \A_1 & \sim (\A^{\rm SM}_{pp \to h_1} \kappa_{3}) \times \frac{ \bar\lambda_{321} \bar\lambda_{211}}{D_3(p)D_2(p')} ,&
    \A_2^{(a)} & \sim( \A^{\rm SM}_{pp \to h_1}   \kappa_{a}) \times \frac{ \bar\lambda_{a111}} {D_a(p)}.
    \label{A1A2}
\end{align}

The inverse propagators are $D_a(p)=p^2-M_a^2+i M_a \Gamma_a $, with $p$ the momentum flowing through the propagator, and $\Gamma_a$ the decay width of $h_a$. On resonance, we have $|p^2-M_a^2| \ll | M_a \Gamma_a|$.

The process $\A_1$ requires at least two extra singlets, which acquire a vev ensuring non-zero mixing angles. The amplitude can be doubly enhanced if both the intermediate $h_3$ and $h_2$ states are resonantly produced. $\A^{(a)}_2$ arises in the TRSM ($a=1,2,3$ if all scalars have a vev), but also in the xSM ($a=1,2$) and the SM ($a=1$). The enhancement is limited to a single resonance of the intermediate singlet $h_a$ state. 

For the process $\A_1$ to dominate triple Higgs boson production, the resonance enhancement should be large enough to overcome the suppression due to the small mixing angles. Experimental data constrains the mixing with the SM Higgs field $|\sin \theta_{12}|,|\sin \theta_{13}| =\O(\eps)$ with $\eps \sim 0.1-0.2$ (see \cite{Papaefstathiou:2020lyp} and \cref{theta}), whereas the singlet-singlet mixing can be arbitrarily large. Expanding in small mixing angles gives
\begin{equation}
    \A_1 = \O(\eps^3), \quad \A_2^{(2)},\A_2^{(3)} = \O(\eps^2), \quad \A_2^{(1)} = \O(\eps^0).
    \label{eps_exp}
\end{equation}
Away from any resonances, the SM-like $\A_2^{(1)}$ process dominates up to small $\O(\eps^2)$ mixing corrections. 

\subsubsection{Benchmark plane ``BP3''}

The collider signatures of the TRSM have been investigated in the literature. In particular, ref.~\cite{Robens:2019kga} identified six benchmark planes (BP) of interest for double and triple Higgs boson production, which were selected to maximize the rates.
This was followed up in~\cite{Papaefstathiou:2020lyp} by a parameter scan of the BP3 plane for triple Higgs boson production, taking all experimental constraints into account. The rates are significantly enhanced compared to the SM, but also much larger than in the xSM. 

The benchmark points for triple Higgs boson production in  the TRSM provided in \cite{Papaefstathiou:2020lyp} all lie in the so-called BP3 plane, characterized by mixing angles and singlet vevs
\begin{equation}
    \theta_{12}=-0.129, \quad
    \theta_{13}=0.226, \quad
    \theta_{23}= -0.899, \quad
    v_2 =140\,{\rm GeV}, \quad v_3 =100\,{\rm GeV}.
    \label{theta}
\end{equation}
The singlet masses are taken in the range $M_2 = [125, 500]\,{\rm GeV}$ and $M_3 = [255, 650]\,{\rm GeV}$.

The leading-order result in the small mixing angles expansion in  \cref{eps_exp} does not yet yield a very accurate approximation since $\eps$ is not that small. However, it gives insight into the qualitative behavior of the amplitudes. At leading order
\begin{align}
   \A_1 \propto \frac{(M_1^2+M_2^2+M_3^2)(M_2^2+2M_1^2)}{D_3(p)D_2(p') v_h v_2 v_3 }
    &\theta_{13} ( v_2 \theta_{13}-v_3 \theta_{12}  )\times \nonumber \\
    &\cos\theta_{23} \sin\theta_{23} (\theta_{13} \sin \theta_{23}-\theta_{12} 
\cos\theta_{23}  ).
\label{A1_approx}
\end{align}
Fixing the angles $\theta_{12},\theta_{13}$, the mixing angle $\theta_{23}$ can be optimized to maximize triple Higgs boson production. For the benchmark parameters \cref{theta}, this gives $|\theta_{23}| \approx -1.1$, not too far off the benchmark value. For the benchmark points $|v_3 \theta_{12}|< |v_2 \theta_{13}|$, and neglecting the $v_3 \theta_{12}$ factor in \cref{A1_approx}, the $v_2$ dependence drops out of the equation. To further maximize the triple Higgs boson production rate one can increase the ratio $M_3/v_3$, or in terms of the couplings in the Lagrangian, increase $\lambda_{33}$. For all benchmark points $\lambda_{33} \sim 5-7$ is chosen.  

Last, but not least, the amplitude is resonantly enhanced if the intermediate scalars are produced on-shell, and the propagators squared can be approximated to have a narrow width. Parametrically, this enhances the cross section by a factor $M_i/\Gamma_i$ for each state $h_i$ above threshold. For both $h_2$ and $h_3$ resonances to occur, the masses should satisfy $M_3 > M_2 + M_1$ and $M_2 > 2 M_1$. See \cref{fig:M2M3plane} for the combination of masses of the BM points that mostly enhance triple Higgs boson production.

\subsection{Perturbativity}
\label{sec:Perturbativity}
As discussed in the previous subsection, the benchmark points all have large quartic couplings -- especially $\lambda_{33}$ -- to enhance the triple Higgs boson signal through resonant production. This prompts us to check the perturbativity of the theory. We will work with the one-loop RGE Renormalization Group Evolution (RGE) equations. Our criterion for convergence of the loop expansion is that the one-loop contribution to a quartic coupling is less than half the tree-level result.\footnote{Our perturbativity cutoff is the scale at which the contributions from two loops and higher become $\mathcal{O}(1)$. To determine more precisely the convergence of the perturbative series requires going to higher loop order, which is beyond the scope of this work, but would be necessary for more accurate phenomenological predictions.} This gives a bound on the quartic couplings, which can be translated into a renormalization scale, beyond which the theory becomes non-perturbative and calculational control is lost. The thus-defined perturbativity cutoff should be well above the scale of the triple Higgs boson production process, set by the masses of the heavier singlets.  The perturbativity cutoff is about a factor of seven below the Landau pole, where the quartic couplings blow up.

The one-loop RGE equations for the quartic couplings are given in \cref{A:RGEs}. The running of the quartic couplings is dominated by the contribution of the quartic couplings themselves, as in the parameter range of interest these are much larger than all SM couplings.  Their RGEs are then of the form
\begin{equation}
    (4\pi)^2\beta_{\lambda_{I}} = a_{I}\lambda_{I}^2 + \sum_{K,J} b_{KJ} \lambda_{K} \lambda_{J}+...
    \label{beta_lambda}
\end{equation}
with $I = ij$ and $i,j=1,2,3$ with $j\leq i$.
Taking the couplings on the right-hand side constant, the solution is
\begin{equation}
    \lambda_{I}(\mu)\approx \lambda_{I}(\mu_0) + \frac{1 }{(4\pi)^2} \(a_{I}\lambda^2_{I}(\mu_0) +\sum_{K,J} b_{KJ} \lambda_{K}(\mu_0) \lambda_{J}(\mu_0) \)\ln\(\frac{\mu}{\mu_0}\),
\end{equation}
which is the one-loop correction to the coupling. All couplings are defined at the scale of the Z-boson mass $\mu_0 = M_Z$.
Demanding the loop contribution to be less than half of the tree-level coupling gives the perturbativity constraint
\begin{equation}
    \(a_{I}\lambda_{I}(\mu_0) +\sum_{K,J} b_{KJ} \frac{\lambda_{K}(\mu_0) \lambda_{J}(\mu_0)}{\lambda_{I}(\mu_0)} \)\ln\(\frac{\mu}{\mu_0}\) < \frac12(4\pi)^2.
    \label{perturbative}
\end{equation}
The theory becomes non-perturbative at the scale $\mu_{\rm pert}$ that saturates the equality above. The theory is already non-perturbative at the electroweak scale if \cref{perturbative} has already been violated for $\ln(\mu/\mu_0) \sim 1$. 
If only one of the couplings is non-perturbatively large, we can drop the $b_{IJ}$-terms, and the constraint becomes $a_I\lambda_I<8\pi^2 $. For the TRSM couplings, the coefficients can be read off from the RGE equations in \cref{RGEs}: $a_{11}=24$, $a_{22,33}=18$ and $a_{12,13,23}=4$, yielding
\begin{align}
    \lambda_{11} < \frac{\pi^2}{3} \approx 3.3, \quad
    \lambda_{22},\lambda_{33} < \frac{4\pi^2}{9}\approx 4.4,\quad
     \lambda_{12},\lambda_{13},\lambda_{23}<2\pi^2 \approx 20.
\label{eq:pert_cond}     
\end{align}
If several couplings are large, the $b_{IJ}$-terms contribute as well, and the constraints generically become stronger. We note that, for the mixed couplings with $a_{12,13,23}$, the stability bounds \cref{eq:bounded} are stronger given the perturbativity bounds on the diagonal self-couplings.
Our perturbativity constraints of \cref{eq:pert_cond} are stronger than those used in~\cite{Robens:2019kga,Papaefstathiou:2020lyp}.

Dropping the mixed $b_{IJ}$-terms, \cref{beta_lambda} can be solved exactly. The solution blows up at the scale $\mu_{\rm pole}$:
\begin{equation}
    a_I\lambda_{I}(\mu_0) \ln(\mu_{\rm pole}/\mu_0) = (4\pi)^2.
\end{equation}
Comparing with \cref{perturbative} we see that the Landau pole and the non-perturbativity scale are related via $\ln(\mu_{\rm pole}/\mu_{\rm pert})=2$ or 
\begin{equation}
    \mu_{\rm pole} \approx 7.4\mu_{\rm pert}.
    \label{pole}
\end{equation}

For all benchmark points in~\cite{Robens:2019kga,Papaefstathiou:2020lyp} the quartic couplings at the scale $\mu_0 =M_Z$ are in the range
\begin{equation}
\label{eq:perturbativity_cond}
    \lambda_{22}(\mu_0) =3.5-4.5,\quad
    \lambda_{33} (\mu_0)= 5.9-6.9,\quad
    \lambda_{23}(\mu_0) =3.9-6.1.
\end{equation}
These values violate \cref{eq:pert_cond}, and clearly the theory is non-perturbative already at the electroweak scale.  The Landau pole for these point is below the TeV scale.

%%%%%%%%%%%%%%%%%%%%%%%%%%%%%%%%%%%%%%%%%%%%%%%%%%%%%%%%%%%%%%%%%%%%%%%%%%%%%%
\section{Updated scan for benchmark points}
\label{sec:scan}

\begin{table}
\begin{center}
\begin{tabular}{ |c|c|c|c|}
\hline
\multicolumn{4}{| c |}{}\\
 \multicolumn{4}{| c |}{\bf{Relevant \texttt{HiggsBounds} Experimental Analyses}}\\
 \multicolumn{4}{| c |}{}\\
\hline
Processes & Experiment & Int. Luminosity & arXiv ref.\\
\hline
$gg\rightarrow S\rightarrow W^+W^-, ZZ$ & ATLAS & 139~fb$^{-1}$ & 2004.14636 \cite{ATLAS:2020fry} \\\hline
$gg\rightarrow S \rightarrow ZZ$ & ATLAS & 139~fb$^{-1}$ & 2009.14791  \cite{ATLAS:2020tlo}\\\hline
$gg \rightarrow S \rightarrow h_1 h_1 \rightarrow(b\bar{b} ) (\tau^+ \tau^-)$ &  CMS & 137~fb$^{-1}$  & 2106.10361 \cite{CMS:2021yci} \\ \hline

$(b\bar{b},\tau^+ \tau^-,W^+W^-,ZZ,\gamma \gamma) (b\bar{b})$ &  & 35.9~fb$^{-1}$ &  1811.09689 \cite{CMS:2018ipl}\\ \hline

$gg \rightarrow S\rightarrow h_1 h_1 \rightarrow $& ATLAS & 36.1~fb$^{-1}$  & 1906.02025 \cite{ATLAS:2019qdc}\\

$(b\bar{b},\tau^+ \tau^-,W^+W^-,\gamma \gamma)^2$ &  &  & \\ \hline

$gg \rightarrow S \rightarrow h_1 h_1 \rightarrow(b\bar{b} ) (\gamma \gamma)$ & ATLAS & 36.1~fb$^{-1}$  & 1807.04873 \cite{ATLAS:2018dpp} \\ \hline

$gg\rightarrow S\rightarrow W^+W^-, ZZ$ & ATLAS & 36.1~fb$^{-1}$ & 1808.02380 \cite{ATLAS:2018sbw}\\\hline
$pp \rightarrow S \rightarrow ZZ$ (incl. VBF) & CMS & 35.9~fb$^{-1}$ & 1804.01939 \cite{CMS:2018amk}\\\hline

$gg \rightarrow S \rightarrow h_1 h_1 \rightarrow(b\bar{b} ) (b\bar{b})$ & CMS & 35.9~fb$^{-1}$  & 1806.03548 \cite{CMS:2018qmt}\\ \hline

$gg \rightarrow S \rightarrow h_1 h_1 \rightarrow(b\bar{b} ) (b\bar{b})$ & ATLAS & 36.1~fb$^{-1}$  & 1806.03548 \cite{CMS:2018qmt}\\ \hline

\hline

\end{tabular}
\caption{The most constraining experimental analyses on new scalar particles in the \texttt{HiggsBounds} library, found during our scan for viable points over the parameter space of the TRSM. In the process description (first column), the particle $S$ denotes either of the $h_2$ or $h_3$ scalars, $S = \{h_2, h_3\}$.}\label{tab:expanalyses}
\end{center}
\end{table}

We have performed a parameter scan over the phase space of the TRSM in order to find new benchmark points for enhanced triple Higgs boson production, for which the theory is perturbative at the EW scale \cref{eq:perturbativity_cond}.  As we saw in~\cref{sec:old_BM}, the points in the BP3 plane defined in~\cite{Robens:2019kga} require large quartic couplings for enhanced triple Higgs boson production, and break the condition given in~\cref{eq:perturbativity_cond}. Hence, in this study, our scan is not limited to this plane, but instead covers the full parameter space. We thus perform a more comprehensive search over the masses and vevs of the singlets, as well as over the corresponding mixing angles. We check that all points are in the perturbative regime, and that the scalar potential is bounded from below as in \cref{eq:bounded}.  

To enforce the fulfillment of the state-of-the-art experimental constraints, we have employed the \texttt{HiggsTools} package~\cite{Boggia:2017hyq,Bahl:2022igd}. The package \texttt{HiggsTools} is a toolbox for comparing a wide class of new physics models to all available experimental results from searches for new scalar particles and measurements of the 125~GeV SM-like Higgs boson at colliders. It is composed of the sub-libraries: \texttt{HiggsSignals} and \texttt{HiggsBounds}. \texttt{HiggsSignals} compares model predictions to a full set of current measurements of the 125~GeV Higgs properties. Using all
available correlation information, it computes and returns a $\chi^2$ value that
quantifies the agreement of the model predictions with the measurements. The viable points in our scans are compatible at 95\% confidence level (C.L.) with respect to the SM (i.e.\ 2$\sigma$).\footnote{The $\chi^2$ is calculated for the SM, and we ask for any ``viable'' point to be within 2$\sigma$ of that value.} \texttt{HiggsBounds} tests
model predictions against a database of experimental results from searches for new
particles. Depending on the parameter point, the most constraining processes are used to compare a theoretical prediction with experimental bounds. The model predictions for all of these
selected limits are then required to lie below the observed limit to obtain an
approximate overall 95\% C.L. exclusion bound. The most constraining experimental bounds applicable are listed in Table~\ref{tab:expanalyses}. 

For the actual scan we have generated $530,000$ random points over the phase space defined by $M_2, M_3, v_2, v_3, \theta_{12}, \theta_{13}, \theta_{23}$.  The ranges considered are as follows:
\begin{align}
&M_2\in [255, 700]~{\rm GeV}, \quad &&M_3\in [350, 900]~{\rm GeV}, \nonumber \\
&v_2 \in [0, 1000]~{\rm GeV}, \quad &&v_3 \in [50, 1000]~{\rm GeV}.
\end{align}
For the mixing angles $\theta_{12}, \theta_{13}, \theta_{23}$  we impose the following limits on the scaling factors~\cite{Papaefstathiou:2019ofh,Robens:2019kga} of~\cref{eq:kappa}:
\begin{equation}
0.95 \leq \kappa_1\leq 1.00,
\quad 
0.0 \leq \kappa_2 \leq 0.25,
\quad
0.0\leq \kappa_3 \leq 0.25.
\end{equation}

To fulfill this task we have used the Python-based scripts which can be found in \cite{gitlabrepoTwoSingletScan}. These programs allow us to calculate all the corresponding self couplings  which are then stored in the  
``run card'' file in the format required by our implementation of the  TRSM for \texttt{MadGraph5\_aMC@NLO} available in \cite{gitlabrepoTRSM}. This file is then provided to \texttt{MadGraph5\_aMC@NLO} to calculate the required cross-section, see below.

Using the scalar couplings we test the 
boundedness from below and the perturbativity of the theory for energy scales ranging from $\mu_0=M_Z\approx 91\,\rm{GeV}$ up to $\mu=900\,\rm{GeV}$. Then, the experimental tests listed in Table~\ref{tab:expanalyses} are evaluated using \texttt{HiggsTools}. Between $20\%$ and $30\%$ of the total number of points passed these initial tests. More concretely, from our $530,000$ random phase space points only $115,056$  passed the first round of theory plus experimental constraints, this is about $22\%$ of the total produced.

For each of the surviving $115,056$ TRSM instances we evaluated the cross section $\sigma_{3 h_1}$ for the triple Higgs boson production $p p \rightarrow h_1 h_1 h_1$ using \cite{gitlabrepoTRSM}. 
We then performed one more test to select the points with enhanced triple Higgs production by demanding a cross-section to be at least 100 times as large as the one in the SM $\sigma^{\rm SM}_{3 h_1}$:
\begin{eqnarray}
\sigma_{3 h_1}> 100\,\sigma^{\rm SM}_{3 h_1},
\label{eq:CScondition}
\end{eqnarray}
and found that only 140 of our TRSM phase space points fulfill this last condition. 

\begin{figure}[h]
\centering
\includegraphics[width=12cm]{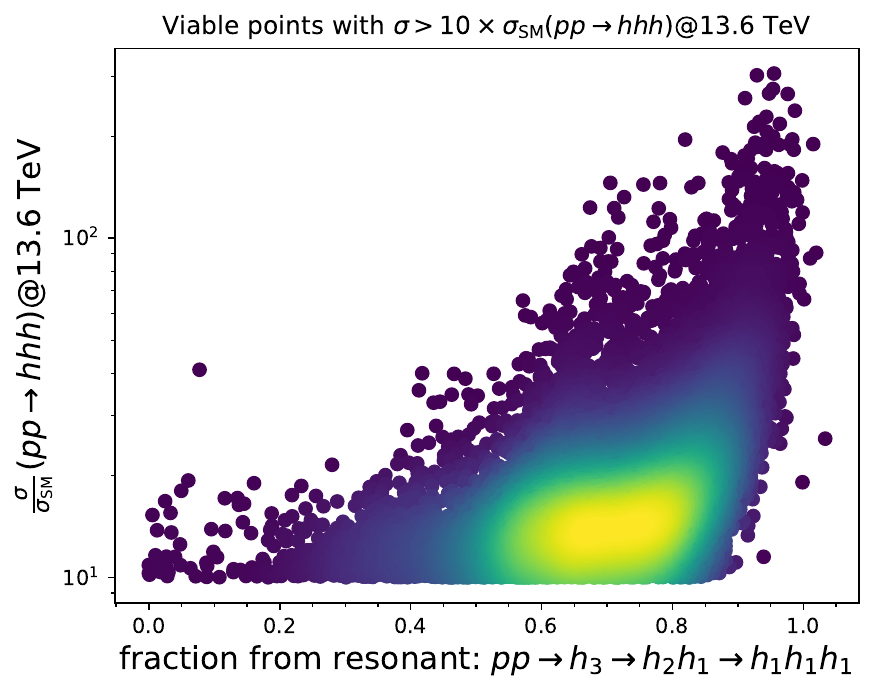}
\caption{ Enhancement of the triple Higgs boson production cross section $\sigma(p p \rightarrow h_1 h_1 h_1)$ at 13.6~TeV, given in terms of multiples of the SM value, and the resonant fraction contribution from $ p p \rightarrow h_3 \rightarrow h_2 h_1 \rightarrow h_1 h_1 h_1$. Only points with a factor 10 enhancement or greater are shown. The density of points increases from the dark blue to yellow shade.}
\label{fig:EnhancementvsresonantFraction}
\end{figure}

\begin{figure}[h]
\centering
\includegraphics[width=12cm]{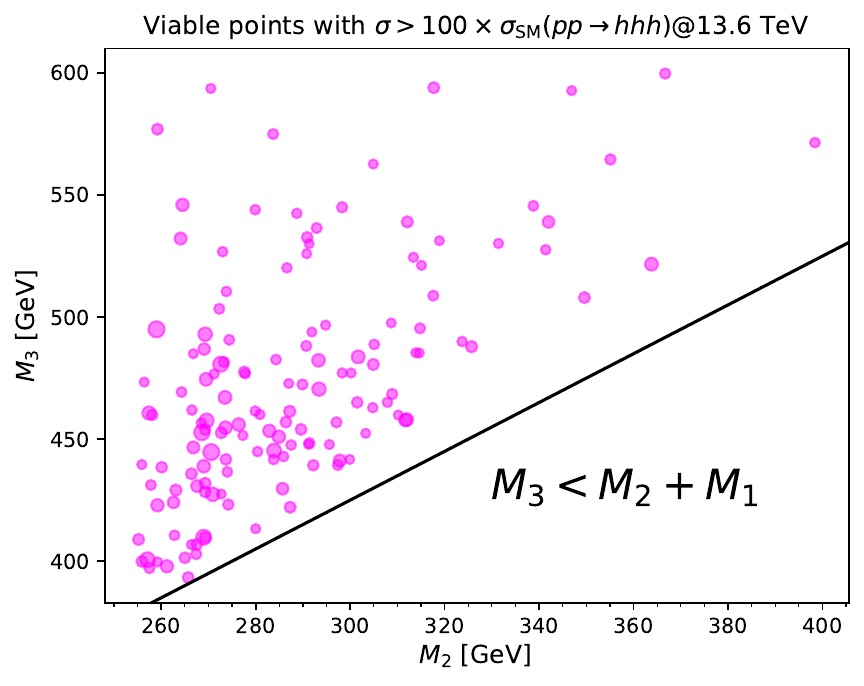}
\caption{Scatter plot of the values of $M_2$ and $M_3$ for the 140 points with triple Higgs boson production cross section over 100 times the SM value. The black solid line denotes the region where double resonant production is kinematically viable, i.e.\ the boundary $M_3 = M_2 + M_1$.}
\label{fig:M2M3plane}
\end{figure}

We have studied the contribution from the channel $p p \rightarrow h_3 \rightarrow h_2 h_1  \rightarrow h_1 h_1 h_1$ to the total cross section  $\sigma_{3 h_1}$. The results are shown in \cref{fig:EnhancementvsresonantFraction}.
The points with the largest enhancement for $\sigma_{3 h_1}$ are also those where the channel $p p \rightarrow h_3 \rightarrow h_2 h_1  \rightarrow h_1 h_1 h_1$ dominates. This is the same effect as reported in~\cite{Papaefstathiou:2020lyp}. It should be noted though, that our present scan is more comprehensive in terms of the coverage of the parameter space, whereas the one in our previous study was restricted to the BP3 plane~\cite{Papaefstathiou:2020lyp}, and in addition, includes points for which the perturbativity constraints are not satisfied. \Cref{fig:M2M3plane} presents a new version of the  distribution of points in the $M_2-M_3$ plane which give enhanced triple Higgs boson production as defined in eq.~\eqref{eq:CScondition}. Finally, we applied the perturbativity tests of \cref{eq:pert_cond}, and determined the scales at which perturbativity gets violated and when we hit the Landau pole.

Moreover, we determined the energy $\mu_{\rm pole}$ at which the RGE leads to divergent values for the self-couplings. The theoretical correlation expected between $\mu_{\rm pert}$ and $\mu_{\rm pole}$
is given in \cref{pole}. This relationship is reasonably obeyed in practice for most of our points, although we found that for few of our benchmark scenarios $\mu_{\rm pole}\sim 8 \mu_{\rm pert}$.
The full set of BM points, including the resonance fraction and the values of $\mu_{\rm pert}$ and $\mu_{\rm pert}/\mu_{\rm pole}$ are provided in the ancillary files. A sample is demonstrated in \cref{tab:benchmark}.

It is interesting to note at this point, that, within the narrow width approximation, we expect the kinematic distributions of the double-resonant process $ p p \rightarrow h_3 \rightarrow h_2 h_1 \rightarrow h_1 h_1 h_1$ to only depend on the masses and widths of the scalar particles. The couplings $\kappa_{123}$ and $\kappa_{112}$, i.e.\ those involving $h_3-h_2-h_1$ and $h_2-h_1-h_1$, respectively, merely rescale the rate for the process. This fact could be exploited in a phenomenological analysis, potentially allowing for a model-independent extraction of constraints. We leave this prospect to future work. 

\begin{table}
\small
\begin{center}
\begin{tabular}{ |c|c|c|c|c|c|c|c|c|c|c|}
\hline
\multicolumn{11}{| c |}{}\\
 \multicolumn{11}{| c |}{\bf{Benchmark points for enhanced triple Higgs production}}\\
 \multicolumn{11}{| c |}{}\\
\hline
&&&&&&&&&  &\\
$M_2$  & $M_3$ & $v_2$ & $v_3$ &  $\theta_{12}$  & $\theta_{13}$ & $\theta_{23}$ & $\frac{\sigma}{\sigma_{SM}}$ & Res. Frac. & $\mu_{\rm pert}$ & $\frac{\mu_{\rm pert}}{\mu_{\rm pole}}$ \\
&&&&&&&&&  &\\
\hline
 259.0& 495.0 & 215.8 & 180.8 & 6.191 & 0.163 & 5.691 & 306.025& 0.955 & 
2.7 $\times 10^2$& 7.3\\
\hline
270.6& 444.7& 122.4 & 847.2 & 0.268 & 0.030 & 0.522 & 302.361 & 
0.929 & 1.8 $\times 10^2$ & 7.3\\
\hline
268.6 & 452.7 & 137.8 & 784.8 & 0.263 & 
0.023 & 0.645 & 275.616 & 0.954&  $2.4 \times 10^2$& 7.3\\
\hline
272.6 & 480.7 & 928.3 & 143.7 & 3.098 & 2.9 & 2.375 & 267.245 & 0.948 & $1.4 \times 10^2$ & 7.2\\
\hline
269.0 & 409.8 & 138.0 & 599.4 & 0.244 & 0.004 & 0.773 & 266.439 &
0.976 & $2.4 \times 10^2$ & 7.2\\
\hline
269.1& 486.9& 227.5& 307.9 & 0.074& 6.149& 2.631&
157.583& 0.956& $4.3 \times 10^2$ & 8.0\\
\hline
259.2& 577.0 & 289.0 & 275.6& 0.137& 6.148& 2.324&
145.470& 0.781& $1.2 \times 10^4$ &
7.2\\
\hline
283.7& 575.0 & 259.4& 330.4& 0.137& 6.152& 2.299&
122.546& 0.779& 3.0 $\times 10^3$ & 7.2\\
\hline
264.3& 469.3& 207.3& 359.5& 0.285& 6.277& 0.692&
119.121& 0.999& 5.4 $\times 10^3$ & 7.3\\
\hline
266.5& 461.9& 653.1& 229.0& 2.889& 3.046& 1.015&
112.794& 0.863& 5.3 $\times 10^4$& 8.0\\
\hline
259.2&399.7& 444.5& 217.0 & 2.917 & 3.046&  1.047&
103.717 & 0.973 & 1.2 $\times 10^5$ & 8.0\\
\hline
\end{tabular}
\caption{ Sample of selected benchmark points obtained during our scan. The first five points correspond to those with the overall highest cross-section $\sigma$ (reported in terms of the ratio $\sigma/\sigma_{\rm SM}$, where $\sigma_{\rm SM}$ is the SM cross section). The remaining six points are selected to illustrate the spectrum of scales $\mu_{\rm pert}$ found during our analysis. The masses $M_2$ , $M_3$, vacuum expectation values $v_2$ and $v_3$ and scale $\mu_{\rm pert}$ are given in GeV.
The 9th column refers to the fractional contribution of the channel $p p \rightarrow h_3 \rightarrow h_2 (\rightarrow h_1 h_1) h_1$ to $\sigma$, it has been estimated as a first approximation by ignoring interference effects. As explained before, here we only present a subset of points for the purposes of illustration, we provide a complete list in the ancillary files.}\label{tab:benchmark}
\end{center}
\end{table}

%%%%%%%%%%%%%%%%%%%%%%%%%%%%%%%%%%%%%%%%%%%%%%%%%%%%%%%%%%%
%%%%%%%%%%%%%%%%%%%%%%%%%%%%%%%%%%%%%%%%%%%%%%%%%%%%%%%%%%%
%%%%%%%%%%%%%%%%%%%%%%%%%%%%%%%%%%%%%%%%%%%%%%%%%%%%%%%%%%%
\section{Thermal history of the TRSM} \label{sec:thermal}

It is well-known that loop corrections from particles in the plasma at finite temperature change the shape of the scalar field potential. As a result, the zero-temperature vacuum, in which all three scalar fields have a vev in our model, is not the vacuum state in the early universe. To end up in today's vacuum, one or more PT(s) had to occur. These PT(s) can be of different types. In an FOPT the vev of one or several fields makes a discontinuous jump, due to a barrier in the potential. Instead, when the fields move continuously from the high-temperature phase to the zero-temperature phase, this is a second-order or cross-over PT. The electroweak PT in the SM is a cross-over \cite{Kajantie:1995kf, Kajantie:1996mn, Kajantie:1996qd, Gurtler:1997hr, Csikor:1998eu, Aoki:1999fi}, but in many BSM models the transition can be of a first-order type, see e.g.~\cite{Caprini:2019egz} and references therein for examples. 

An FOPT is phenomenologically very interesting for two reasons: it can source a gravitational wave signal possibly observable by LISA, and it can provide the out-of-equilibrium dynamics necessary for electroweak baryogenesis. Both require the FOPT to be strong; for baryogenesis in particular, the SM Higgs vev should obtain a large value in the transition and
\begin{equation}
    \frac{\phi_1}{T} \gtrsim 1,
\end{equation}
at the time of the transition \cite{Kuzmin:1985mm, Manton:1983nd, Klinkhamer:1984di}. Here the $\phi_i$ denote the temperature-dependent classical field values with $\phi_i(T=0) =v_i$.

It would be interesting to correlate the parameter space of enhanced triple Higgs production in the TRSM (and enhanced collider phenomenology in general) with the parameters for which a (sufficiently strong) FOPT can occur.
For this, we study the effective potential at finite temperature using the framework of thermal field theory. As has become clear in recent years, accurate quantitative statements about PTs require careful many-loop computations~\cite{Croon:2020cgk, Niemi:2021qvp, Gould:2023jbz, Kierkla:2023von}. However, to determine \emph{if} an FOPT can occur in the TRSM, as a first approximation we can rely on much simpler expressions for the finite-temperature potential, that can be obtained from standard references, such as~\cite{Dolan:1973qd, Weinberg:1974hy}, or textbooks~\cite{Kapusta:2006pm, Bellac:2011kqa}. 

%%%%%%%%%%%%%%%%%%%%%%%%%%%%%%%%%%%%%%%%%%%%%%%%%%%%%%%%%%%%%%%%%%
\subsection{The finite temperature effective potential}

The one-loop TRSM effective potential at finite temperature is:
\begin{equation}
    V_T(\phi_i,T) = V(\phi_i) + V_{\rm CW}(\phi_i) + V_{\rm c.t.}(\phi_i)
    + V_{T,\, \rm{ 1-loop}}(\phi_i, T),\label{eq:V1loop}
\end{equation}
with $\phi_i$ the field values defined in \cref{eq:fields} (with $\phi_i=v_i$ in the vacuum today).
$V(\phi_i)$ is the tree-level potential of \cref{L_TRSM}, $V_{\rm CW}$ the standard zero-temperature one-loop `Coleman-Weinberg' potential and $V_{\rm c.t.}$ the corresponding counterterms. The temperature-corrections are captured by $V_{T,\, \rm{ 1-loop}}$, which is given by
\begin{equation}
    V_{T, {\rm 1-loop}}(\phi, T) = \frac{T^4}{2\pi^2} \left[\sum_{\alpha = \Phi_i, W,Z} n_\alpha J_B[m_\alpha^2(\phi_i)/T^2] +n_t J_F[m_t^2(\phi_i)/T^2] \right],\label{eq:VT1loop}
\end{equation}
where the masses $m^2_\alpha$ follow from the tree-level potential.

The first sum gives the contribution of all bosonic degrees of freedom and runs over the scalars $\Phi_i$ and the $W$ and $Z$ bosons. We neglect the Goldstone bosons, as their contribution is challenging to treat, since their mass squared can be negative, and their contribution to the PT is typically subdominant~\cite{Cline:2009sn}.
The second term arises from the fermionic degrees of freedom; we only include the top quark here, as the contributions of the lighter fermions can be neglected.  $n_\alpha$ and $n_t$ denote the number of degrees of freedom for each species. The functions $J_{B,F}$ are given by
\begin{equation}
    J_{B,F}(y^2) = \int_0^\infty dk k^2 \log{[1 \mp \exp{[-\sqrt{k^2 + y^2}]}]},
\end{equation}
with the minus sign for bosons and the plus sign for fermions.

At temperatures large compared to the mass, the functions $J_{B,F}$ can be expanded in $m^2_\alpha(\phi_i)/T^2$ as
\begin{align}
    J_B(m_\alpha^2/T^2) &= -\frac{\pi^4}{45} + \frac{\pi^2}{24} \frac{m_\alpha^2}{T^2} -\frac{\pi}{6} \frac{m_\alpha^3}{T^3} - \frac{1}{32}\frac{m_\alpha^4}{T^4} \left(\log{\frac{m_\alpha^2}{16\pi^2 T^2}} - \frac 3 2 + 2\gamma_E \right) \cdots\,\nn , \\
    J_F(m_\alpha^2/T^2) &= \frac{7\pi^4}{360} - \frac{\pi^2}{24} \frac{m_\alpha^2}{T^2} - \frac{1}{32}\frac{m_\alpha^4}{T^4} \left(\log{\frac{m_\alpha^2}{\pi^2 T^2}} - \frac 3 2 + 2\gamma_E \right) \cdots \, ,\label{eq:JB/F}
\end{align}
where $\gamma_E$ is the Euler-Mascheroni constant, and we have discarded an infinite sum of terms of higher order in $m_\alpha^2/T^2$. The $m_\alpha$-independent term contributes to the total energy of the universe, but is not relevant for the PT, so we ignore it. The virtue of the high-temperature expansion is that it allows us to obtain some analytical understanding of the effect of the separate terms in the effective potential. We will use this in section~\ref{sec:LO-PT} to investigate the possibility of PTs at leading order.
However, the effective potential in the high-temperature expansion has to be treated with care. In particular, fields with masses $m^2_\alpha(\phi_i) \gtrsim T^2$ get Boltzmann-suppressed.
The full one-loop thermal function of eq.~\eqref{eq:VT1loop} takes this suppression into account, but the potential corresponding to eq.~\eqref{eq:JB/F} does not. In section~\ref{sec:Vcubic}, we will study the evolution of the fields as given by eq.~\eqref{eq:VT1loop}, without the high-temperature expansion.

%%%%%%%%%%%%%%%%%%%%%%%%%%%%%%%%%%%%%%%%%%%%%%%%%%%%%%%%%%%%%%%%%%%%%%%%%%%%
\subsection{Leading order phase transitions}\label{sec:LO-PT}

We will first consider the potential with only the leading-order thermal mass corrections in \cref{eq:JB/F} included. The phase transition dynamics crucially depend on the zero-temperature minimum of the potential, which we will refer to as `the vacuum'. We will show in this subsection that if all scalars obtain a vev -- the case of interest for triple Higgs boson phenomenology -- then none of the phase transitions is first-order. Instead, if one or both of the singlets remain at zero field value, a first-order phase transition (FOPT) is possible; for completeness, we discuss this case in \cref{sec:other_min}.

As can be seen from \cref{eq:VT1loop,eq:JB/F},
at leading order in $m_\alpha/T$, the effect of the thermal plasma is to give the scalar fields an effective \emph{thermal mass}. Explicitly, the masses (in the symmetric phase) are \cite{Fonseca:2020vke, Ekstedt:2022bff}:
\begin{align}
    m_{1}^2(T) &= -\mu_1^2 + \frac{T^2}{48} \left(3 g_1^2 + 9 g_2^2 + 2(6 y_t^2 + 12\lambda_1 + \lambda_{12} + \lambda_{13}) \right), \nonumber \\
    m_{2}^2 (T) &= -\mu_2^2 + \frac{T^2}{24}\left(4 \lambda_{12} + \lambda_{23} + 6 \lambda_2 \right), \label{mass_thermal} \\ 
    m_{3}^2 (T) &= -\mu_3^2 + \frac{T^2}{24}\left(4 \lambda_{13} + \lambda_{23} + 6 \lambda_3 \right), \nonumber
\end{align}
resulting in an \emph{effective} finite-temperature potential:
\begin{equation}
    V_{\rm eff, LO} (\phi_i, T) = \frac12 \sum_i m^2_i(T) \phi_i^2 + \frac14 \sum_{i \leq j}
\lambda_{ij} \phi_i^2 \phi_j^2 .\label{eq:VeffTtree}
\end{equation}
From these expressions, we can immediately understand why one or more PT(s) occur as the universe cools down. At high temperature the thermal corrections can make some or all of the masses positive, and the global minimum at early times differs from the zero-temperature vacuum.  

Note that for the updated benchmark points discussed in Sec.~\ref{sec:scan}, the global minimum does not always lie at the origin at high temperatures in the perturbative regime. Firstly, the scale where the masses become positive may lie above the perturbativity cut-off. Secondly, depending on the couplings, some of the thermal masses in \cref{mass_thermal} can be negative, leading to a period of `symmetry non-restoration' \cite{Weinberg:1974hy, Mohapatra:1979vr, Dvali:1995cc, Meade:2018saz, Baldes:2018nel, Matsedonskyi:2020mlz}. In the TRSM at least the Higgs thermal mass is always positive, because of the large top quark Yukawa contribution and the boundedness of the negative couplings \cref{eq:bounded}, and thus the electroweak symmetry always gets restored at high temperature. For concreteness, we will consider the case where all thermal masses are positive at temperatures below the perturbativity cutoff, and track the field dynamics as the universe cools down and all fields obtain a vev. If one or multiple thermal masses are already negative at the perturbativity scale, the analysis still applies, but starting from the point where there are already some non-zero vevs. As we  will show, one can never have two minima (and thus a FOPT)
at the same time, irrespective of the order in which extrema appear. Our results thus do {\it not} depend on the assumption that at high temperature all masses are positive and the symmetry is fully restored.

The PT dynamics in the TRSM mirrors that of the xSM, analyzed in e.g.~\cite{Espinosa:2007qk, Profumo:2007wc, Cline:2009sn, Espinosa:2011ax, Curtin:2014jma, Alves:2018jsw, Gould:2019qek}. In the well-studied two-step scenario \cite{Land:1992sm}, only the Higgs field has a non-zero vev in the vacuum $(h,s) = (v_h,0)$. Depending on the parameters, it is possible to have a two-step transition, where the first transition $(0,0) \to (0,s)$ is continuous, but the 2nd step $(0,s) \to (h,0)$ is an FOPT. However, if both fields have a vev in the zero-temperature vacuum $(h,s) = (v_h,v_s)$, at leading order all possible transitions are second order/cross over as they are always from a saddle point or maximum to a minimum. This has also been demonstrated in \cite{Espinosa:2011ax, Ghorbani:2020xqv}.

Our findings can be understood intuitively. As the temperature drops, the thermal mass of one of the fields, say $x_1$, first becomes negative. A new minimum appears starting at the origin and moving continuously to larger field values along the $x_1$-direction as  $m_1^2(T)$ decreases, while the origin becomes a saddle. The $x_1$ field thus obtains a continuously growing vev in a smooth non-first-order transition. At later times, a second field, say $x_2$, will develop a negative mass (the mass matrix is evaluated at a nonzero $x_1$-vev). The new minimum first appears at the same point as the old minimum, and will then continuously move away along the $ x_2$ direction, while the previous minimum becomes a saddle. This continues until all three fields have a vev. The full series of transitions
\begin{equation}
    (0,0,0) \to (x_1,0,0) \to (x_1,x_2,0) \to (x_1,x_2,x_3)
\end{equation}
is continuous, as each time the new minimum appears, it emerges from the previous one which in turn becomes a saddle. This sequence of events relies on all fields only appearing quadratically in the potential, as a consequence of gauge symmetry and the $\mathbb{Z}_2$-symmetries in \cref{eq:VeffTtree}.

In the rest of this subsection, we will prove the above statements more rigorously. We will find the extrema of the potential, and determine when they are real (i.e.\ when they exist) and when they are minima. This analysis reveals an important property: in the limit of an extremum becoming a real solution, it emerges from another solution that already existed before, and the corresponding transition is continuous.

\subsubsection{Extrema}
We start from the finite-temperature potential \cref{eq:VeffTtree}, but  denote the scalars as $x_i$ instead of $\phi_i$, where $x_i$ is the $i$th field to obtain a vev:
\begin{equation}
     V(x_1, x_2, x_3) = \frac12\sum_i m_i^2 x_i^2 + \frac14\sum_{i,j} c_{ij} x_i^2 x_j^2,
\end{equation}
with $c_{ij}=c_{ji}$. The normalization of $c_{ij}$ is slightly different than before, explicitly $\lambda_{ii} =c_{ii}$ and $\lambda_{ij} =2c_{ij}$ for the diagonal and non-diagonal couplings respectively; this convention simplifies the expressions in this subsection. Here and in the following repeated indices do \emph{not} imply summation. The extrema are the points for which all derivatives of the potential vanish:
\begin{gather}
  \forall k, \quad    \partial_{x_k} V = m_k^2 x_k +  \sum_{i} c_{ik} x_i^2 x_k = 0.
     \label{ext}
\end{gather}
%where we introduced the notation $V_k = \partial_{x_k} V$ and, for later use, $V_{kl} = \partial_{x_k}\partial_{x_l} V$. 
Because of the $\mathbb{Z}_2$ symmetry, we can focus on the extrema with non-negative vev without loss of generality.
The extremum condition then has two solutions %for each field 
for
 each $k$:
\begin{gather}
    x_k=0 \quad \quad \lor \quad   m_k^2+  \sum_{i} c_{ik} x_i^2 = 0. \label{eq:Extremum}
\end{gather}
There are then four qualitatively different extrema, shown in the left panel of \cref{fig:transitions}, which we refer to as the `origin', and as the `axial', `planar', and `bulk' extremum respectively. 
\begin{itemize}
    \item Origin: $\vec x_0 \equiv (0,0,0)$.
\item Axial extremum $\vec x_1 \equiv (x_1,0,0)$ with
\begin{equation}  x_1= \sqrt{-m_1^2/c_{11}}. 
\label{axial}
\end{equation}
 \item Planar extremum $\vec x_{12} \equiv (x_1,x_2,0)$ with
    \begin{equation}
        x_1= \sqrt{\frac{c_{12}m_2^2 - c_{22}m_1^2}{c_{11}c_{22} - c_{12}^2}}, \quad
        x_2=\sqrt{\frac{c_{12}m_1^2 - c_{11}m_2^2}{c_{11}c_{22} - c_{12}^2}}.
        \label{planar}
    \end{equation}
     \item Bulk extremum $\vec x_{123} \equiv (x_1,x_2,x_3)$ with
     \begin{align}
    x_1 &= \frac{{\sqrt{(c_{23}^2-c_{22} c_{33}) m_1^2 + (c_{12} c_{33}-c_{13} c_{23}) m_2^2 + (c_{13} c_{22}-c_{12} c_{23}) m_3^2}}}{{\sqrt{D}}},
  \nn  \\
    x_2 &= \frac{{\sqrt{(c_{12} c_{33}-c_{13} c_{23} ) m_1^2 + (c_{13}^2-c_{11} c_{33} ) m_2^2 + (c_{11} c_{23}-c_{12} c_{13} ) m_3^2}}}{{\sqrt{D}}},    
   \nn  \\
    x_3 & = \frac{{\sqrt{( c_{13} c_{22} - c_{12} c_{23}) m_1^2 + (c_{11} c_{23} - c_{12} c_{13}) m_{2}^{2} +  (c_{12}^{2} - c_{11} c_{22}) m_{3}^{2}}}}{{\sqrt{D}}}, 
    \label{bulk}
\end{align}
where
\begin{equation}
D = c_{11} c_{22} c_{33} + 2 c_{12} c_{13} c_{23} - c_{13}^2 c_{22} - c_{11} c_{23}^2 - c_{12}^2 c_{33},
\end{equation}
is the determinant of $c_{ij}$.
\end{itemize}
The other axial ($\vec x_2, \vec x_3$) and planar ($\vec x_{13}, \vec x_{23}$) solutions are obtained by relabeling the field indices, as the potential is invariant under these permutations.

The extremum is a minimum if the eigenvalues of the Hessian of the potential $h_{kl}$, i.e. the mass matrix,
evaluated at the extremum are all positive, with
\begin{equation}
    h_{kl}(x_1, x_2, x_3) \equiv \partial_{x_k} \partial_{x_l} V(x_1, x_2, x_3) = ( m_k^2 + \sum_{i} c_{ik} x_i^2) \delta_{kl} + 2 c_{kl} x_k x_l. \label{hes}
\end{equation}
The Hessian is block diagonal for the axial, planar, and bulk extrema. We can analyze the upper-left block -- the  $1\times 1$ block for $\vec x_1$, the $2\times 2$ block for $\vec x_{12}$ and $3\times 3$-matrix for $\vec x_{123}$ -- by rescaling the fields $x_i^2 \to y_i$ with $y_i>0$, and calculating the Hessian for the $y_i$-coordinates. For the directions of non-zero field value $x_i>0$, i.e.\ for the blocks of the Hessian mentioned above, the rescaling is a monotonically increasing function and will not affect the sign of the eigenvalues. The rescaled hessian is 
\begin{equation}
    \bar h_{kl}(y_1, y_2, y_3)=\frac12 c_{kl}.
    \label{hes_rescale}
\end{equation}
We demand that  $\vec x_{123}$  is today's vacuum. The eigenvalues of the rescaled Hessian should then be positive. Sylvester's criterion, stating that a square Hermitian matrix is positive definite if \textit{and only if} all the leading principal minors are positive, then gives 
\begin{equation}
c_{ii} > 0, \quad \& \quad  C_{ij} \equiv c_{ii}c_{jj}-c_{ij}^2 > 0,  \quad \&\quad D > 0,
\label{cond_min}
\end{equation}
for $i\neq j$.

%%%%%%%%%%%%%%%%%%%%%%%%%%%%%%%%%%%%%%%%%%%%%%%%%%%%%%%%%%%%%%%%%%%%%
\subsubsection{Phase transition dynamics} {

Let us now look at the extrema in more detail, and discuss the possible PTs between them.
We start at high temperature with all thermally corrected masses $m_1^2$, $m_2^2$, and $m_3^2$ positive and will analyze how the bulk minimum, today's vacuum, is reached.
Without loss of generality, we assume that $x_1$ first becomes non-zero, then $x_2$, and finally $x_3$. This means $m_1^2$ becomes negative first, followed by  the mass of $x_2$ evaluated at $\vec x_1$, followed by the mass of $x_3$ evaluated at $\vec x_{12}$. 
As mentioned above, if at the perturbativity scale already one or more of the masses are negative, the dynamics discussed in this subsection equally apply if we just start at a later point in the field trajectory from the origin to the final minimum with non-zero vevs for all fields.

\paragraph{Origin}
The Hessian of the potential at the origin $\vec x_0 =(0,0,0)$ is
\begin{equation}
    h_{kl}(\vec x_0) = m_k^2 \delta_{kl}. \label{Hes0}
\end{equation}
At high temperature, when all quadratic mass terms are positive $m_k^2>0$, the origin is a minimum of the potential. The origin is then also the only extremum, as the axial, planar and bulk extrema \cref{axial,planar,bulk} do not have a real solution, i.e., do not yet exist.  

\paragraph{Axial extremum}
As $m_1^2$ becomes negative, a new extremum emerges at $\vec x_1$ given in \cref{axial}. The Hessian \cref{hes} at this axial extremum is diagonal 
\begin{equation}
   h(\vec x_1) = {\rm diag}(-2 m_1^2,\,m_2^2 - m_1^2 c_{12}/c_{11},\,m_3^2 - m_1^2 c_{13}/c_{11} ).
   \label{hes_axial}
\end{equation}
With the sign flip of $m_1^2$ the origin has become a saddle, while the first eigenvalue of the Hessian at $\vec x_1$ is positive. With $m_2^2,m_3^2$ still positive and large enough all other eigenvalues are positive as well, and $\vec x_1$ is (temporarily) the global minimum of the potential.
The axial extremum and the extremum at the origin coincide at the moment the former becomes real, that is
\begin{equation}
    \lim_{m^2_1 \to 0^-} \vec x_1 =  \vec x_0. \label{x1x0}
\end{equation}
As the temperature drops further, $m_1^2$  becomes more negative and the extrema separate continuously from each other.  The transition from $\vec x_0$ to $\vec x_1$ is thus a continuous PT from a saddle point to a minimum, not an FOPT.

Finally, we note that there can never be two axial minima at the same time. For $m_i^2,m_j^2<0$ the $\vec x_i,\vec x_j$ extrema both exist. A necessary condition for both axial extrema to be a minimum is that  $h_{jj}(\vec x_i)>0$ and $h_{ii}(\vec x_j)>0$, which can only be satisfied for $c_{ij}>0$, which then implies  $C_{ij}<0$, which is at odds with our choice of vacuum \cref{cond_min}. Therefore, there cannot be an FOPT between different axial minima.

\paragraph{Planar extremum}
The planar extremum comes into existence when
$\vec x_{12}$ in \cref{planar} becomes real, which requires
\begin{equation}
  c_{12} m_1^2 - c_{11} m_2^2= -h_{22}(\vec x_1) c_{11}   >0  , \quad \& \quad
   c_{12}m_2^2 - c_{22} m_1^2=-h_{11}(\vec x_2) c_{22} >0,
    \label{range_planar}
\end{equation}
where we have used $C_{12} >0$.
For $c_{12}>0$ both conditions can only be satisfied for negative mass $m_2^2<0$, while in the opposite case $c_{12}<0$ the $\vec x_{12}$ extremum already emerges when $m_2^2$ is still positive, provided it is small enough $m_2^2 <| m_1^2c_{12}|/c_{11}$. 

The Hessian at $\vec x_{12}$ is block diagonal. The eigenvalues of the upper-left $2\times 2$ block are positive if the upper-left $2\times 2$ block of the rescaled Hessian \cref{hes_rescale} is positive, which is ensured by \cref{cond_min}. Thus   $\vec x_{12}$ is a minimum if the third eigenvalue is positive as well, which reads
\begin{align}
    h_{33} (\vec x_{12}) &=  m_3^2 - \frac{c_{13} c_{22} h_{11}(\vec x_2) + c_{23} c_{11}h_{22}(\vec x_1)}{ C_{12}}
    = h_{33}(\vec x_1) +\frac{\rho}{C_{12}} h_{22}(\vec x_1)     >0,
    \label{h33}
\end{align}
with ${\rho} \equiv c_{12}c_{13} - c_{23}c_{11}$.
This is satisfied for large enough $m_3^2$. 

Just as the axial minimum emerged from the origin, the planar minimum coincides with the axial extremum when it forms
\begin{equation}
    \lim_{ h_{22}(\vec x_1) \to 0^-} \vec x_{12} = \vec x_{1}.
     \label{x12x2}
\end{equation}
As the temperature drops, $h_{22}(\vec x_1)$ gets more negative and the extrema separate from each other continuously. The first condition in \cref{range_planar} implies that the moment $\vec x_{12}$ becomes real, the axial extremum $\vec x_1$ turns into a saddle, as $h_{22}(\vec x_1)$ -- the mass of $x_2$ evaluated at $\vec x_1$ -- becomes negative.  Likewise, the second condition implies $\vec x_2$ is a saddle. The origin also still is a saddle, as its Hessian has at least one negative eigenvalue.
Until the bulk minimum becomes real, the planar extremum $\vec x_{12}$ is the global minimum of the potential.   The transition from $\vec x_1$ to $\vec x_{12}$ is a continuous PT from a saddle point to a minimum.  

It remains to be shown that there can be no FOPT between $\vec x_{12}$ and $\vec x_3$, and between $\vec x_{12}$ and $\vec x_{13}$;  these possibilities do not arise in the set-up with a single singlet such as the xSM. We use a proof by contradiction to show that such FOPTs do not arise.

Let's start with planar to axial (or axial to planar) transitions $\vec x_{12} \leftrightarrow \vec x_{3} $. Assume that $\vec x_{12}$ and $ \vec x_{3} $ are both real minima. This implies positive eigenvalues of their respective Hessian, $h_{33}(\vec x_{12})>0$ and 
$h_{ii}(\vec x_{3})>0$ for $i=1,2,3$, and in addition $m_3^2<0$ for the $\vec x_{3} $ extremum to exist. It is possible to find an algebraic relation connecting these quantities using \cref{hes_axial} (by permutation of indices) and \cref{h33}:
\begin{equation}
    -m_3^2\frac{D}{c_{33}} + C_{12} h_{33}(\vec x_{12}) = {\sigma}  h_{11}(\vec x_{3}) + {\rho} h_{22}(\vec x_{3}),
    \label{relation_x3}
\end{equation}
with ${\sigma} \equiv c_{12}c_{23} - c_{13}c_{22}$ and  as before ${\rho} = c_{12}c_{13} - c_{23}c_{11}$.  Since $D,C_{12}>0$ by our choice of vacuum \cref{cond_min}, the left-hand side of this equation is positive. 
As we will now show, our starting assumption that $\vec x_{12}, \vec x_{3} $ are minima implies that $\rho,\sigma<0$ and the right-hand side is negative -- we run into a contradiction. The proof depends on the sign of $c_{12}$:
\begin{enumerate}
\item
If $c_{12}>0$,  all thermal masses have to be negative for $\vec x_{12}$ and $\vec x_3$ has to exist, and $c_{13}<0$ for $h_{11}(\vec x_3)$ has to be positive. Positive Hessian factors $h_{11}(\vec x_{3}),\, h_{22}(\vec x_{3})$ then dictate that $h_{33}(\vec x_{1}),\, h_{33}(\vec x_{2}) < 0$ are negative for $C_{13}>0$.
Now combining the right-most expression in \cref{h33} with the first condition in \cref{range_planar} gives $\rho < 0$. By permutation of indices in \cref{h33} we also conclude that $\sigma < 0$. 

\item  If $c_{12}<0$, $m_1^2,m_3^2$ have to be negative and $m_2^2$ positive for $\vec x_{12}$ and $\vec x_3$ has to exist, and $ c_{13}>0$ for $h_{11}(\mathbf{x_3})>0$.  The sign of $c_{23}$ is still free. If $c_{23}>0$, it is easily checked that $\rho,\sigma<0$ as all individual terms are negative. If $c_{23}<0$,  demanding $h_{11}(\vec x_{3}),h_{22}(\vec x_{3})$ both positive limits $m_2^2 > |m_3^2 c_{23}|/c_{33} > |m_1^2 c_{23}|/c_{13}$; combining with \cref{range_planar} then gives  $\rho,\sigma<0$.
\end{enumerate}
We thus conclude that in all cases $\vec x_3$ and $\vec x_{12}$ cannot be minima at the same time, as this violates \cref{relation_x3}, and there is thus no FOPT between them.

Finally, we consider a possible planar to planar transition $\vec x_{12} \to \vec x_{13}$. For both extrema to be a minimum, $h_{33}(\vec x_{12}) >0$ in \cref{h33}, and a similar relation follows by permutation of the indices for $h_{22}(\vec x_{13}) >0$. Keeping in mind \cref{range_planar}} both can only be satisfied simultaneously if
\begin{equation}
\rho^2 - C_{12}C_{13} = - c_{11} D   >0.
\label{plan-plan}
\end{equation}
However, the positivity of the determinant $D>0$ forbids this possibility. There are thus no planar to planar FOPT transitions.

\paragraph{Bulk extremum.}

For the transition to the bulk minimum, the reasoning and results are similar to the preceding ones. The $\vec x_{123}$ solution becomes real for $m_3^2$ small enough, and it emerges from the planar extremum:
\begin{equation}
    \lim_{h_{33}(\vec x_{12}) \to 0^-} \vec x_{123} = \vec x_{12}.
\end{equation}
That is, in the limit of $\vec x_{123}$ becoming real, it emerges from $\vec x_{12}$, which therefore needs to be real as well. As we saw before, if $\vec x_{12}$ is a real solution, then the axial extrema and the origin are not minima.  Demanding that the $x_3$ component of $\vec x_{123}$ in \cref{bulk} to be real gives $h_{33}(\vec x_{12})<0$, indicating that the planar extremum $\vec x_{12}$ now is a saddle as well. Using the symmetry of the potential under permutations of the field order, we can draw the same conclusion for the other two components of $\vec x_{123}$. As $h_{33}(\vec x_{12})$ becomes more negative, the bulk minimum separates from the planar extremum, and becomes the global minimum of the potential, as we have demanded from the start, cf. \cref{cond_min}.

\begin{figure}
    \centering
    \includegraphics[width=0.45\linewidth]
    {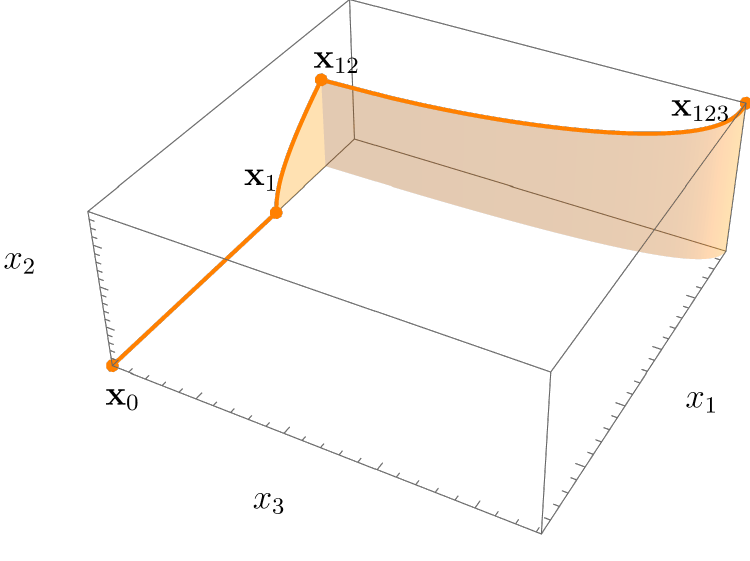}
    \includegraphics[width=0.45\linewidth]
    {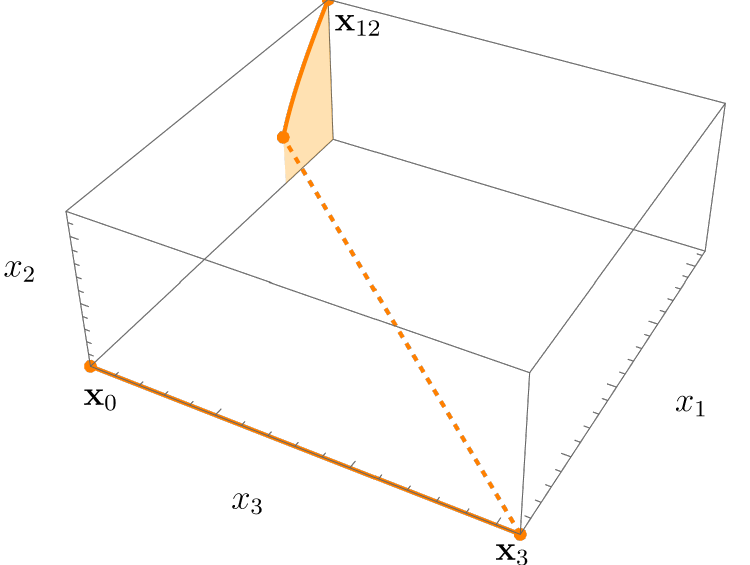}
    \caption{Visualization of possible phase transitions. To the left is the case of the low-temperature vev at $\vec x_{123}$. To the right is the case of low-temperature vev at $\vec x_{12}$}
    \label{fig:transitions}
\end{figure}

We have already seen that two axial, two planar, and an axial and planar extrema can never be minima at the same time and that the origin is not a minimum once the other extrema emerge. Now our final analysis of the bulk minimum reveals that once it comes into existence, the planar (but also all other) extrema are not minima. Hence, as depicted in the left panel of \cref{fig:transitions}, all possible transitions are smooth and continuous, and no FOPT arises.
This conclusion on the absence of an FOPT is based on the bulk minimum being the zero temperature vacuum of theory, and on only including the leading-order thermal corrections in \cref{eq:VeffTtree}. In the next subsections, we will relax those assumptions.

%%%%%%%%%%%%%%%%%%%%%%%%%%%%%%%%%%%%%%%%%%%%%%%%%%%%%%%%%%%%%%%%%%%%%
\subsubsection{Axial or planar minimum as today's vacuum}
\label{sec:other_min}
In this section we drop our assumption that the zero-temperature vacuum is the bulk extremum, and focus instead on the possibility of an axial or planar minimum today; this corresponds to respectively none or only one of the additional singlets having a vev at zero temperature. We do not attempt a full analysis of the parameter space, but we limit the discussion to point out the conditions that allow for an FOPT. For the sake of readability, we will denote today's vacuum by simply `vacuum' in this section.

\paragraph{Axial vacuum}
We take $\vec x_1$ as the vacuum. This scenario is not favorable for multi-Higgs boson production enhancement, but we will see that it accommodates an FOPT. Let us consider the transition
\begin{equation}
\vec x_2 \to \vec x_1.
\label{ax_ax}
\end{equation}
With all $C_{ij}<0$ negative, none of the planar extrema are minima as the appropriate block of the rescaled Hessian \cref{hes_rescale} for the $\vec x_{ij}$ solution has determinant $C_{ij}<0$, and thus negative eigenvalues. The bulk extremum is neither the vacuum, as \cref{cond_min} no longer holds. Lastly, we require that $V(\vec x_1)< V(\vec x_2)$ at zero temperature. The mass $m_2^2$ goes negative before $m_1^2$, and the extremum $\vec x_2$ emerges first.  The setup is qualitatively similar to the last step of a two-step transition in the xSM. Indeed, the additional $x_3$-scalar of the TRSM does not obtain a vev and is just a spectator during the PT.  

As is known from the xSM literature, the axial-axial transition can be first-order. For this to happen both axial extrema should be minima, and their respective Hessian $h(\vec x_1)$ and $h(\vec x_2)$ in \cref{hes_axial} should have positive eigenvalues. Since $C_{12}<0$ -- unlike the bulk vacuum -- this is now possible, provided that 
\begin{equation}
    m_1^2,\, m_2^2 < 0, \quad C_{12}<0, \quad m_3^2 > m_1^2 c_{13}/c_{11},\,m_2^2 c_{23}/c_{22}.
\end{equation}
With negative masses the planar extremum $\vec x_{12}$ also exists, but is a saddle point in the $(x_1, x_2)$ plane. It follows that there can be an FOPT between the two axial minima through the barrier formed by the potential near $\vec x_{12}$. 

\paragraph{Planar vacuum}
Consider now that the planar extremum $\vec x_{12}$ is the vacuum, which requires $C_{12}>0$.
From the collider point of view, this represents mixing between the Higgs field and only one added scalar, leading to a phenomenology similar to the xSM and a possible enhancement of di-Higgs boson production. For the phase transition, however, there are two distinct possibilities for an FOPT that are unique to the TRSM and have no equivalent in the xSM.

We first discuss the axial to planar transition. As shown in the previous subsection, the planar extremum and the axial extrema from which it emerges -- for example $\vec x_{12}$ and $\vec x_1,\, \vec x_2$ -- cannot be minima at the same time. This leaves us with the transition
\begin{equation}
    \vec x_3 \to \vec x_{12}.
\end{equation}
If the bulk extremum is the vacuum, $\vec x_3$ and $\vec x_{12}$ cannot be minima at the same time, since the left- and right-hand side of \cref{relation_x3} would then have a different sign. For $c_{12}>0$ this conclusion was based on $D,C_{13},C_{23}$ all positive, while for $c_{12}>0$ we only used a positive determinant $D$. However, with the planar vacuum, we can flip the sign of the left-hand side of \cref{relation_x3} by taking the determinant negative (for $c_{12} >0$ there is also the possibility to flip the sign of the right-hand side with $C_{13}, C_{23}$ negative). The determinant of the $c_{ij}$ matrix can be expressed as
\begin{equation}
    D = c_{33} C_{12} + c_{13}\sigma + c_{23}\rho < 0.
    \label{Dneg}
\end{equation}
For definiteness, we take all $C_{ij}>0$ and $c_{12}>0$ and all masses negative, the same parameters as in case 1 discussed below \cref{relation_x3}; as we saw previously, $\vec x_3$ and $\vec x_{12}$ can then be both minima if in addition $c_{13}>0$ and $c_{23}>0$, which gave  $\rho,\sigma<0$. The determinant can be negative and \cref{relation_x3} can be satisfied as well for
\begin{equation}
    0 < h_{33}(\vec x_{12}) < \frac{D m_3^2}{c_{33} C_{12}}.
\end{equation}

When $\vec x_3$ is a minimum, the other planar extrema $\vec x_{13}, \vec x_{23}$ do not exist yet. The bulk extremum \cref{bulk} is however real since all masses $m_i^2$ are negative. $C_{12}>0$ and $D<0$ imply that the first two eigenvalues of the scaled Hessian in \cref{hes_rescale} are positive, while the third is negative, meaning that the bulk extremum is a saddle point. We conclude that there can be an FOPT between $\vec x_3$ and $\vec x_{12}$ with the barrier formed by the potential near the bulk saddle. An illustration of this case is on the right-hand panel of \cref{fig:transitions}. As mentioned before, this case is particularly interesting for LHC phenomenology as a low-temperature vacuum where the Higgs field and one other field have a non-zero vev can lead to an observed enhancement of di-Higgs boson production.

Finally, we point out the possibility of a planar to planar FOPT, of the type
\begin{equation}
    \vec x_{13} \to \vec x_{12}.
\end{equation}
For the bulk vacuum, we saw that the condition \cref{plan-plan} prohibits both planar extrema to be a minimum. However, for the planar vacuum, we can again avoid this by taking the determinant \cref{Dneg} negative. For the same parameters as above, the bulk extremum is a saddle which can create
a barrier between the two minima at $\vec x_{12}$ and $\vec x_{13}$, allowing for an FOPT between them.

\subsection{Phase transitions in the full one-loop effective potential}\label{sec:Vcubic}
Let us now go beyond the leading order finite-temperature effect in the potential. In the high-temperature expansion, the next contribution comes from the term in eq.~\eqref{eq:JB/F} proportional to $m_\alpha^3$ in $J_B$. A barrier can thus be generated by the interplay between positive quadratic and quartic terms, and a negative cubic term. This scenario is typically referred to as a `radiatively-generated barrier', and it is the relevant mechanism for e.g.\ the one-step PT in the xSM. Because of the rather large couplings of our BM points, and the mass hierarchies existing at zero temperature, we have to be careful not to include contributions of fields with $m(\phi_i) \gtrsim T$ in eq.~\eqref{eq:JB/F}. We therefore work instead with the full one-loop thermal function of eq.~\eqref{eq:VT1loop}, in which heavy fields become automatically Boltzmann-suppressed.

Note that the cubic terms correspond to IR-divergent diagrams \cite{Dolan:1973qd,Kirzhnits:1976ts}. This IR-divergence is associated with loop contributions from massless modes, and it gets cured by a resummation of these masses - this procedure is often referred to as Arnold-Espinosa or daisy resummation \cite{Arnold:1992rz}. Here, we will refrain from including this correction, as its usual implementation critically relies on the high-temperature expansion. Introducing the resummation term for fields that should be Boltzmann-suppressed, would result in an unphysical leftover dependence on the heavy field. 

In order to look for FOPT(s), we numerically determine the values of the scalar fields in the global minimum of the potential of eq.~\eqref{eq:V1loop} as a function of the temperature, using the function \texttt{NMinimize} of \texttt{Mathematica}. If we find any discontinuities in the values of the scalar fields, we can interpret that as an indication that an FOPT might occur. We use running couplings, which we evaluate at the scale $\mu_T \equiv 1.25 \pi T$, corresponding to the scale that we use in the Coleman-Weinberg potential. This implies that, as we increase $T$, we approach the scales $\mu_{\rm pert}$ and $\mu_{\rm pole}$. For $\mu_T \gtrsim \mu_{\rm pert}$ we can not expect our analysis to be reliable, and the PT would have to be studied on the lattice.\footnote{Strictly speaking, PT(s) with couplings satisfying our perturbativity bounds still become non-perturbative in the IR. It has, however, been demonstrated that a perturbative study with enough orders included can reproduce the lattice calculations~\cite{Kajantie:2002wa,Ekstedt:2022zro, Gould:2023ovu}.} We will therefore focus on points where the PT(s) take place below this scale.\footnote{It should be noted that even for PT(s) at temperatures with $\mu_T \gtrsim \mu_{\rm pert}$ we still do not find indications of an FOPT}

A complication in the evaluation of the potential is that the squared masses entering in eq.~\eqref{eq:VT1loop} and in the CW term are not necessarily positive. When taken at face value, these terms lead to imaginary contributions in the potential. These imaginary parts have been demonstrated to be associated with the decay rate of the quantum state \cite{Weinberg:1987vp}, and it has also been argued that the imaginary contributions of the CW-potential largely get canceled out by the imaginary contributions of eq.~\eqref{eq:VT1loop} and the daisy-resummation term \cite{Delaunay:2007wb}, which resums the massless modes. 
Since we can not consistently include the daisy-resummation term, we inevitably have to deal with a complex potential. 
In the literature, different strategies are chosen to tackle this complication, for example, taking the real part of the potential \cite{Delaunay:2007wb} or the absolute values of the masses \cite{Athron:2022jyi}. Unfortunately, these treatments lack a robust theoretical motivation and they lead to unphysical kinks in the potential (see Figure 1 of~\cite{Profumo:2007wc}), and consequently in the evolution of the background field.
Imaginary parts in the potential can be avoided altogether by adhering to a strict perturbative expansion around the leading order minimum of the effective potential \cite{Fukuda:1975di, Nielsen:1975fs, Gould:2021ccf, Ekstedt:2022zro, Gould:2023ovu}. This approach requires a careful tracking of the temperature- and field-dependent masses of the fields, and the construction of an appropriate effective field theory for the (light) fields taking part in the PT. This more precise, but also more elaborate approach would be necessary to obtain reliable predictions when we expect an FOPT to occur. 
As we are currently still investigating \emph{whether} an FOPT occurs, we will resort to the somewhat \emph{ad hoc} approach of plugging in the absolute values of the mass into eq.~\eqref{eq:V1loop}. We find this approach to be numerically better behaved than taking the real part, as the evolution of the Higgs field, which is our main interest, does not suffer from large numerical (unphysical) jumps. 

\begin{figure}
    \centering
\includegraphics[width = 0.55\textwidth]{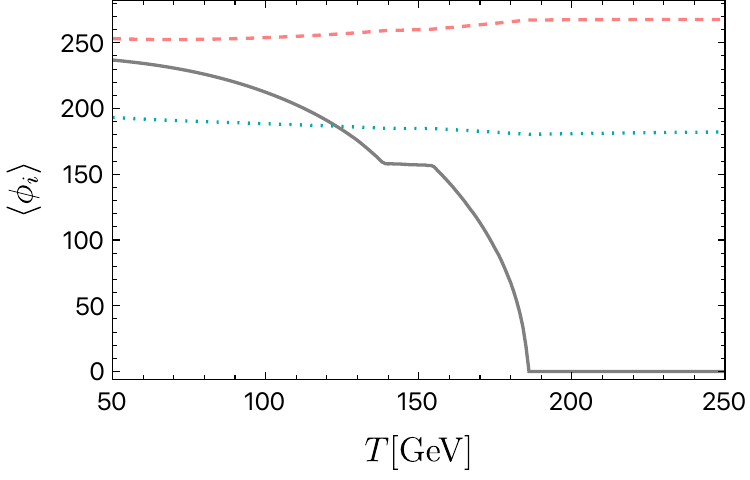}
    \caption{Evolution of the field expectation values in the minimum of the potential for a representative third BM point in Table~\ref{tab:benchmark}. The Higgs field is represented by gray solid, $\phi_2$ by dashed pink, and $\phi_3$ by dotted cyan.}
    \label{fig:BM15}
\end{figure}

Let us first focus on the PT of the Higgs field. By inspecting all 140 BM points, and focusing on temperatures where the couplings are perturbative, we find no hints of an FOPT. 
A representative evolution of the vevs of the fields as a function of the temperature is illustrated by figure~\ref{fig:BM15}. The vevs are found by using the function \texttt{NMinimize} in \texttt{Mathematica}.
The fields $s_1$ and $s_2$ obtain a vacuum expectation value first (because their thermal mass becomes negative first, or because the thermal corrections become Boltzmann-suppressed), which makes them too massive to contribute significantly to the PT of the Higgs field. For the BM point displayed in Figure~\ref{fig:BM15} for example, at $T = 200 \,{\rm GeV}$ the two eigenvalues of the mass matrix are $605 \,{\rm GeV}$ and $472 \,{\rm GeV}$, resulting in Boltzmann-suppression of their contributions. Also, for non-zero values of the Higgs field, these mass eigenvalues are roughly constant, so these fields can not contribute to the cubic barrier of the Higgs field.
The Higgs PT then proceeds in a manner very close to the case in the SM, although the existence of the other two fields affects its quartic coupling, which would even reduce the strength of the PT -- if it was first-order-- which is inversely proportional to the Higgs quartic coupling, see e.g.~\cite{Cline:2006ts, Quiros:2007zz, Morrissey:2012db}.
In all cases, we see that the Higgs field obtains a vev at a temperature of $T \sim 200\, {\rm GeV}$, and that the other fields then have a vev close to their zero-temperature values.
We thus find no indication that the TRSM can explain electroweak baryogenesis in the regime where triple Higgs boson production is significantly enhanced. 

Our results are in line with, and complement the results of~\cite{Cline:2009sn}, where the EWPT in the singlet-majoron model was studied. In an extensive parameter scan, the authors find that, for the parameter space corresponding to SM with a single additional singlet, there are no strong first-order phase transitions with a zero-temperature non-zero singlet vev. Note that ref.\cite{Carena:2019une} \emph{does} find a region with an FOPT and a non-zero singlet vev at zero temperature. Note that the singlet is light in this study, which is therefore not in tension with our results.

One can still hope that the PT of the new scalars is first-order, possibly resulting in an observable GW signal. We don't find any indications of this in any of the BM points, although the absolute value in the mass term results in some numerical jumps in the values of the fields, so our numerical results are not entirely conclusive. The absence of such a PT at leading order, demonstrated in section \ref{sec:LO-PT} does not make such a transition plausible.

\section{Conclusions}\label{sec:conclusions}

We set out to connect Higgs phenomenology at colliders with the electroweak phase transition in the early universe. For definiteness, we choose the two-real singlet extension of the SM (TRSM) with $\mathbb{Z}_2$ symmetry as an example model. Compared to the SM, the TRSM enhances multi-Higgs boson production, a process which is under investigation in the ongoing Run 3 and the upcoming high-luminosity run of the LHC. Moreover, extensions of the Higgs sector similar to the TRSM are known to generate the cosmological first-order phase transition required to produce the observed matter-antimatter asymmetry of the universe. 

What sets the TRSM apart from SM extensions with a single additional field is that triple Higgs production can be enhanced by a double resonance. This requires both new scalars to have a vev. Rates can be further boosted if some of the quartic singlet couplings are large, see~\cref{A1A2}. Demanding that the couplings are still perturbative at least up to the energy scale of the heaviest singlet state \cref{perturbative}, results in stricter bounds than previously used in the TRSM literature \cref{eq:pert_cond}. This has motivated us to find a new set of $140$ benchmark points that enhance the cross-section of triple Higgs boson production within current experimental constraints and theoretical bounds. The new benchmark points can be found in the ancillary files and a subset is presented in \cref{tab:benchmark}.  In the future, it might be interesting to update the TRSM benchmark points of \cite{Robens:2019kga} for di-Higgs boson production and triple-Higgs productions with different mass hierarchies as well, given the perturbativity bounds introduced in this paper.

Previous studies of the electroweak phase transition in BSM models such as the xSM have shown that the chance of the transition being first order -- in the SM it is a crossover -- is better for large scalar couplings. A first-order phase transition (FOPT) arises if there is a sufficiently large barrier between separate minima of the potential at some moment in the cosmological history. Such a barrier (which may already exist at zero temperature) can be generated and deformed due to high-temperature effects, giving rise to an FOPT. These effects may come from the temperature-dependence of the leading order thermal masses, or via radiative corrections.
We have shown analytically in \cref{sec:LO-PT}, that the first option does not generate the needed barrier if both added scalars attain a non-zero vev at low temperatures. As for a radiatively induced barrier, no analytical proof is available, but our numerical study shows that no FOPT occurs for any of the BM points. Therefore, the $\mathbb{Z}_2$ TRSM parameter space for triple Higgs boson production and that of FOPT appear to be mutually exclusive. To have an FOPT, at least one scalar should have a zero field value at low temperatures. In this scenario, resonant di-Higgs boson production is still possible, enhancing the di-Higgs boson cross-section at the LHC, as one would find within the xSM \cite{No:2013wsa, Chen:2014ask, Dawson:2015haa,Li:2019tfd}.

Lastly, we point out two ways to achieve both an FOPT and enhanced triple Higgs boson production. The first is to add additional scalar(s) to the $\mathbb{Z}_2$ symmetric TRSM. Then, two scalars can have a non-zero vev at low temperatures, contributing to triple Higgs resonance production, and the other(s) can have a zero vev at low temperatures, facilitating an FOPT. Another way is to allow $\mathbb{Z}_2$ symmetry-breaking terms to the TRSM. It has been shown that the xSM without $\mathbb{Z}_2$-symmetry can exhibit an FOPT while enhancing di-Higgs boson production~\cite{Kotwal:2016tex, Alves:2018oct, Papaefstathiou:2021glr}, so similarly, breaking the $\mathbb{Z}_2$ symmetry in TRSM would be expected to accommodate the possibility of both an FOPT and triple Higgs boson production enhancement. We leave both of these possibilities for future work.

\acknowledgments

We would like to thank Oliver Gould, Wolfgang Kilian, Carlo Pandini, Philipp Schicho and Tuomas Tenkanen for enlightening discussions.  A.P.\ acknowledges support by the National Science Foundation under Grant No.\ PHY 2210161. This project has received support from the European Union’s Horizon
2020 research and innovation programme under the Marie Sklodowska-Curie grant agreement No
945422 [G.T-X.]. 
This research has been
supported by the Deutsche Forschungsgemeinschaft (DFG, German Research Foundation)
under grant 396021762 - TRR 257 ``Particle Physics Phenomenology after the Higgs Dis-
covery''.
JvdV is supported by the Dutch Research Council (NWO), under project  number VI.Veni.212.133.

\appendix

%%%%%%%%%%%%%%%%%%%%%%%%%%%%%%%%%%%%%%%%%%%%%%%%%%%%%%%%%%%%%%%
%%%%%%%%%%%%%%%%%%%%%%%%%%%%%%%%%%%%%%%%%%%%%%%%%%%%%%%%%%%%%%%
%%%%%%%%%%%%%%%%%%%%%%%%%%%%%%%%%%%%%%%%%%%%%%%%%%%%%%%%%%%%%%%
\section{Useful relations for the TRSM}\label{A:trmsrelations}

%%%%%%%%%%%%%%%%%%%%%%%%%%%%%%%%%%%%%%%%%%%%%%%%%%%%%%%%%%%%%%%
\subsection{Boundedness of the potential}\
\label{A:boundedness}

The potential is bounded from below if the following conditions on the couplings are satisfied~\cite{Papaefstathiou:2020lyp}:
\begin{equation}\label{eq:bounded}
    \begin{aligned}
        \lambda_{11}, \lambda_{22}, \lambda_{33} &>0, \\
        \bar \lambda_{12} \equiv \lambda_{12} + 2\sqrt{ \lambda_{11} \lambda_{22}} &>0, \\
        \bar \lambda_{13} \equiv \lambda_{13} + 2\sqrt{ \lambda_{11} \lambda_{33}} &>0, \\
        \bar \lambda_{23} \equiv \lambda_{23} + 2\sqrt{ \lambda_{22} \lambda_{33}} &>0, \\
        \sqrt{\lambda_{11}} \lambda_{23} + \sqrt{\lambda_{22}} \lambda_{13} + \sqrt{\lambda_{33}} \lambda_{12} + \sqrt{\lambda_{11}\lambda_{22}\lambda_{33}} + \sqrt{\bar\lambda_{12}\bar\lambda_{13}\bar\lambda_{23}}&>0.
    \end{aligned}
\end{equation}
%%%%%%%%%%%%%%%%%%%%%%%%%%%%%%%%%%%%%%%%%%%%%%%%%%%%%%%%%%%%%%%
\subsection{Mass and flavor basis}
\label{A:basis}

The mass and flavor states are related by
\begin{equation}
    \begin{pmatrix}
        h_1 \\h_2\\h_3
    \end{pmatrix} = R \begin{pmatrix}
        \phi_1 \\\phi_2\\\phi_3
    \end{pmatrix}\;,
    \label{rotation}
\end{equation}
with the rotation matrix $R$ given by
\begin{equation}\label{eq:Rmat}
    R = \begin{pmatrix}
        c_{12} c_{13}             & -s_{12} c_{13}             & -s_{13}     \\
        s_{12} c_{23}-c_{12} s_{13} s_{23} & c_{12} c_{23}+ s_{12} s_{13} s_{23} & -c_{13} s_{23} \\
        c_{12} s_{13} c_3+s_{12} s_{23} & c_{12} s_{23}-s_{12} s_{13} c_{23}  & c_{13} c_{23}
    \end{pmatrix}\;,
\end{equation}
with $c_{ij} = \cos \theta_{ij}$ and $s_{ij} = \sin \theta_{ij}$.

The  $\lambda_{ij}$ and the $\mu_i$  parameters in the scalar potential in \cref{L_TRSM} are given in terms of the angles, masses and vacuum expectation values by
\begin{equation}
    \begin{aligned}
        \lambda_{ii} & =\frac{1}{2 v^2_i}\sum_{k}  M_k^2 R^2_{ki},&\lambda_{ij} & =\frac{1}{v_i v_j}\sum_{k}  M_k^2 R_{ki} R_{kj},&
        - \mu_i &= v^2_i \lambda_{ii} + \frac{1}{2}\sum_{k\neq i}v^2_{k}\lambda_{ik},    \end{aligned}\label{eq:TRSMlams}
\end{equation}
where the mixed couplings  $\lambda_{ij}$ are defined for $i<j$.

%%%%%%%%%%%%%%%%%%%%%%%%%%%%%%%%%%%%%%%%%%%%%%%%%%%%%%%%%%%%%%%
\subsection{RGEs}
\label{A:RGEs}

The one-loop RGEs for the quartic couplings are
\begin{align}
    (4\pi)^2\beta_{\lambda_{11}} &= 24\lambda_{11}^2+\frac{\lambda_{22}^2}{2}+\frac{\lambda_{33}^2}{2}+\frac{3 }{8}g_1^4+\frac{9}{8}g_2^4+\frac{3}{4} g_1^2 g_2^2-6 y_t^4-4\lambda_{11}\gamma_{\Phi_1}, \nn \\
     (4\pi)^2\beta_{\lambda_{22}} &= 18\lambda_{22}^2+2\lambda_{12}^2+\frac{\lambda_{23}^2}{2}, \nn \\
     (4\pi)^2\beta_{\lambda_{33}} &= 18\lambda_{33}^2+2\lambda_{13}^2+\frac{\lambda_{23}^2}{2}, \nn \\
     (4\pi)^2\beta_{\lambda_{12}} &= 4\lambda_{12}^2+
    12\lambda_{12} \lambda_{11}+6\lambda_{12}\lambda_{22}+ \lambda_{13}\lambda_{23} - 2\lambda_{12}\gamma_{\Phi_1},\nn \\
     (4\pi)^2\beta_{\lambda_{13}} &= 4\lambda_{13}^2+
    12\lambda_{13} \lambda_{11}+6\lambda_{13}\lambda_{33}+ \lambda_{12}\lambda_{23} - 2\lambda_{13}\gamma_{\Phi_1},\nn \\ (4\pi)^2\beta_{\lambda_{23}} &= 4\lambda_{23}^2+
    6\lambda_{23} \lambda_{22}+6\lambda_{23}\lambda_{33}+ 4\lambda_{12}\lambda_{13}, 
\label{RGEs}
\end{align}
with $\beta_\lambda = \mu\partial \lambda/\partial  \mu$ and $\gamma_{\Phi_1} = \left(\frac{3 g_1^2}{4}+\frac{9 g_2^2}{4}-3 y_t^2\right)$. The running of the gauge couplings and the top quark is as in the SM
\begin{align}
  (4\pi)^2\beta_{g_i} &= b_i g_i^3,\nn \\
  (4\pi)^2\beta_{y_t}&=\frac{9}{2} y_t^3-
y_t(\frac23 g_1^2+9g_3^2) -y_t \gamma_{\Phi_1}, 
\end{align}
with $b_i=(41/6,-19/6,-7)$ for $i=1,2,3$.

%%%%%%%%%%%%%%%%%%%%%%%%%%%%%%%%%%%%%%%%%%%%%%%%%%%%%%%%%%%%%%%
%\subsection{Finite temperature potential}
%\label{A:VT}

% Bibliography
\clearpage
%% [A] Recommended: using JHEP.bst file
\bibliographystyle{JHEP}
\bibliography{biblio.bib}

\end{document}